\documentclass{aa}  

\usepackage{hyperref}
\usepackage[dvipsnames]{xcolor} 
\hypersetup{
    colorlinks=true,
    linkcolor=blue,
    filecolor=magenta,
    citecolor=teal,      
    urlcolor=cyan,
    }

\usepackage{float}
\usepackage{graphicx}
\usepackage{txfonts}
\usepackage{lipsum}
\usepackage{subcaption}         
\usepackage{lscape}             
\usepackage{placeins}           

\newcommand{\msun}{\hbox{M$_{\odot}$}}

\begin{document}

   \title{FASTAR - I. Continuous and differentiable evolutionary stellar population models}


   \author{Ignacio Martín-Navarro\inst{1,2}, Alexandre Vazdekis\inst{1,2}, Luis Peralta de Arriba\inst{3}, Isaac Alonso Asensio\inst{1,2}, Eirini Angeloudi\inst{1,2}, Patricia Iglesias Navarro\inst{1,2}, Francesco La Barbera\inst{4}, Katja Fahrion\inst{5}, Tereza Jerabkova\inst{6}, Michael A. Beasley \inst{1,2,7}, Jes\'us Falc\'on-Barroso\inst{1,2}, Marc Huertas-Company\inst{1,2,8,9}, Sebasti\'an F. S\'anchez\inst{10,1,2}, Prashin Jethwa\inst{4} }

   \institute{
Instituto de Astrof\'{\i}sica de Canarias,c/ V\'{\i}a L\'actea s/n, E38205 - La Laguna, Tenerife, Spain\\
\email{imartin@iac.es}
\and
Departamento de Astrof\'isica, Universidad de La Laguna, E-38205 La Laguna, Tenerife, Spain
\and 
Departamento de Inteligencia Artificial, Universidad Nacional de Educaci\'on a Distancia (UNED), Calle Juan del Rosal 16, E-28040 Madrid, Spain
\and
INAF-Osservatorio Astronomico di Capodimonte, sal. Moiariello 16, Napoli 80131, Italy
\and
Department of Astrophysics, University of Vienna, Türkenschanzstraße 17, 1180 Wien, Austria
\and 
Department of Theoretical Physics and Astrophysics, Faculty of Science, Masaryk University, Kotlářská 2, Brno 611 37, Czech Republic
\and
Centre for Astrophysics and Supercomputing, Swinburne University, John Street, Hawthorn, VIC 3122, Australia
\and
Observatoire de Paris, LERMA, PSL University, 61 avenue de l'Observatoire, F-75014 Paris, France
\and 
Universit\'e Paris-Cit\'e, 5 Rue Thomas Mann, 75014 Paris, France
\and
Instituto de Astronom\'ia, Universidad Nacional Aut\'onoma de M\'exico, A.P. 106, Ensenada 22800, BC, M\'exico
}

   \date{Received; accepted}
   \titlerunning{FASTAR I}
   \authorrunning{Martín-Navarro et al.}

  \abstract
   {The development of evolutionary stellar population models is central to interpreting observations of galaxies in terms of astrophysical quantities. Stellar population models must therefore be both accurate and compatible with inversion algorithms in order to extract meaningful information from the observed data. Here we present FASTAR, a fully differentiable stellar population synthesis code. Contrary to traditional, grid-based single stellar population models, FASTAR can be continuously evaluated at any age (between 20 Myr and 14 Gyr), metallicity ($-2.5 \le$ [M/H] $\le +0.3$), and initial mass function (IMF). Changes in the IMF parameterization are straightforward, allowing for consistent conversions of colors, magnitudes, and mass-to-light ratios, as well as the synthesis of models under the assumption of arbitrary IMF functional forms. FASTAR provides detailed spectroscopic predictions over the MILES wavelength range ($3,540$--$7,400$ $\AA$) as well as more coarsely sampled spectral energy distributions across a wider 2,000-to-12,000\ $\AA$, which can be directly convolved with any arbitrary set of photometric filters. FASTAR performs at the same level of state-of-the-art simple stellar population models benchmarked against observations of globular clusters and high signal-to-noise spectra of early-type galaxies, but it is faster, lighter, and more flexible. Moreover, its differentiable nature allows for a quantitative understanding of model behavior and uncertainties, as well as a natural framework for gradient descent inference algorithms.
   }


\maketitle
\nolinenumbers 

\section{Introduction}

This is the first paper of a series where we introduce the FASTAR models. Here we explain the basic ingredients and synthesis principles in comparison with state-of-the-art evolutionary stellar population models. Upcoming papers will describe the synthesis of semi-resolved populations and some immediate applications of the unique features of FASTAR.

The synthesis of evolutionary stellar population models combines stellar templates with knowledge on stellar evolution theory to generate so-called simple stellar population (SSP) predictions. The goal of these SSP models is to represent the flux emitted by an integrated stellar population (i.e., where individual stars cannot be resolved) of a given age and chemical composition. The foundational steps of Beatrice Tinsley \citep[e.g.,][]{Tinsley68,Tinsley72,Tinsley76, Tinsley80} have been refined and improved by subsequent generations of modelers to an extent where SSP models have become virtually ubiquitous for extragalactic studies \citep[e.g.,][]{Worthey94,Leitherer99,bc03,TMB:03,Maraston05,Schiavon07,miles,Conroy12}.

The adoption of these models is not coincidental but has been driven by their undeniable success in reproducing the spectro-photometric properties of galaxies. Evolutionary stellar population models have been instrumental in describing now well-established concepts of galaxy formation, from the downsizing of early-type galaxies \citep[e.g.,][]{Peletier89,Worthey92,vazdekis:97,Trager00,Thomas05} to the apparent excess of low-mass stars in the cores of most massive galaxies \citep{vandokkum,spiniello12,labarbera,MN15a,Parikh18}. In combination with complementary information, evolutionary SSP model predictions have helped shape our current view of the Universe \citep[e.g.,][]{Kauffmann03,Gallazzi05,Schawinski07,Peng10,Behroozi13,arjen14,Madau14,Maiolino19,Sebastian20}. 

The calculation of SSP models relies on three main assumptions. First, the total flux emitted by an integrated population is assumed to be the sum of the fluxes of its individual stars. Consequently, SSP predictions are based on sets of stellar templates, which can be either theoretical \citep[e.g.,][]{Gustafsson08,Coelho14,Irigoyen21,Meszaros24} or empirical \citep[e.g.,][]{cat,Pat06,Chen14,Villaume17,Mastar}. Second, the number of stars formed and present in an integrated population is determined by the initial mass function (IMF). The IMF is typically assumed to be universal and to follow the characteristic shape measured in the solar neighborhood \citep{Salp:55,Scalo98,kroupa,Chabrier}, although variable IMF models are also becoming widespread. Finally, the defining assumption of evolutionary SSP models is that, for a given chemical composition, the combination of stars used in the calculations is determined by means of theoretical isochrones \citep{basti1,basti2,Girardi00,Choi16,Hidalgo18}. In other words, SSP models exploit our knowledge on stellar evolution theory to generate accurate spectro-photometric predictions.

With these assumptions, SSPs are delivered to the community as grids of pre-computed models covering a range of ages and metallicities defined by the underlying isochrones. Variations in the IMF are also allowed in some of these models through a discrete sampling of the adopted functional form. In recent years, state-of-the-art SSP models have been extended toward broader spectral coverage \citep[e.g.,][]{Beny,Vazdekis16,Villaume17,Verro22}, spanning from the ultraviolet to the infrared. They also now encompass wider ranges of ages \citep{Levesque13,youngMILES,Hawcroft25} and chemical compositions, including variations in individual elemental abundances \citep{Conroy18,Knowles23}. In addition, increasingly large and complete stellar libraries have been assembled \citep{Mastar, Maraston20}, alongside more refined evolutionary tracks and predictions for non-canonical stellar components \citep{HP13,Byrne22,Lu24}.

In parallel, spectral inversion algorithms have also become progressively more sophisticated thanks to formal Bayesian approaches \citep[e.g.,][]{Acquaviva11,Chevallard16,Carnall18,MN19,prospector,Wang24} and the adoption of machine-learning-based tools \citep[e.g.,][]{Alsing20,Hahn22,Melchior23,Liang23,Angthopo24,Hunt24,Patricia24,Patricia25,Alsing24}. These new ways of analyzing the spectral properties of galaxies allow a better characterization of the expected uncertainties and degeneracies, optimizing the amount of information retrieved from the observed data. Moreover, they enable the rapid and systematic analysis of large spectro-photometric datasets, which would otherwise be unfeasible \citep{Hahn22,Patricia24,Zacharegkas25}.

This evolution in inversion algorithms has also highlighted some of the inherent limitations of state-of-the-art SSP models. In particular, the computation of new models is relatively expensive, which hampers their adaptability to changes in the underlying assumptions (e.g. variations in the IMF functional form, isochrones etc.). Furthermore, the size of the model grids scales exponentially with the number of free parameters, effectively limiting either the sampling or the number of variables (age, metallicity, elemental abundances, and IMF parameterization). Likewise, a grid-based set of SSP models limits the synthesis of continuous and differentiable predictions, which can only be achieved at a less fundamental level \citep[e.g. through the use of star formation histories;][]{Hearin23}. Finally, in most cases \citep[but see][]{Robotham25,Bellstedt25} the community of users does not have access to the actual computation of the SSP models. Given their fundamental and widespread use, increasing the transparency of these computations remains a desirable improvement.

With FASTAR we propose an alternative approach to exploiting the advantages of SSP modeling by releasing the first differentiable code to synthesize evolutionary stellar population models, suitable for both spectroscopic and photometric applications. The layout of this paper is as follows: In Sect.~\ref{sec:ingredients} we present the ingredients that go into the model synthesis. In Sect.~\ref{sec:resu} we describe the outputs of FASTAR, which are then compared to observed data in Sect.~\ref{sec:data}. Finally, Sect.~\ref{sec:summary} provides a summary of the main results, advantages, and prospects of the proposed approach.

\section{Ingredients} \label{sec:ingredients}

The synthesis of the FASTAR models follows the standard, isochrone integration-based approach \citep[see e.g.,][]{miles,Conroy13}. For a given age and metallicity, the flux emitted by an integrated stellar population is given by the following equation:

\begin{equation}\label{eq:1}
F_\lambda\bigl(\mathrm{age},[\mathrm{M}/\mathrm{H}]\bigr)
= \int_{m_{\mathrm{low}}}^{m_{\mathrm{high}}(\mathrm{age})}
S_\lambda^{M_V}\!\left(m \mid \mathrm{age},[\mathrm{M}/\mathrm{H}]\right)
\,\chi(m)\,\mathrm{d}m,
\end{equation}

\noindent
where $S_\lambda^{M_V}\!\left(m \mid \mathrm{age},[\mathrm{M}/\mathrm{H}]\right)$ is, for a given age and metallicity, the spectrum of a star with mass $m$. The mapping between $m$ and the atmospheric parameters effective temperature $T_{\mathrm{eff}}$, surface gravity $\log g$, and metallicity is determined by stellar evolution theory, i.e., by the assumed set of isochrones. While the low-mass cutoff $m_{\mathrm{low}}$ remains fixed typically at around the hydrogen-burning mass limit (0.1 $\mathrm{M}_{\odot}$ in our case, determined by the mass range predicted by the isochrones), the upper mass limit $m_{\mathrm{high}}(age)$ depends on the age of the population through the isochrones. Finally, these stellar spectra are weighted by the number of stars, i.e., the IMF $\chi(m)$.

Equation~\ref{eq:1} is rather generic and describes the basis of most evolutionary synthesis codes. To actually synthesize the models, some additional choices can be made. In particular, in FASTAR we follow the approach of the MILES models \citep{miles,Vazdekis15}, where stellar spectra are not normalized according to their bolometric luminosities but based on the expected V-band absolute magnitudes ($S_\lambda^{M_V}$). This alleviates potential systematics related to the calculation of the bolometric luminosity of the stellar templates but requires the adoption of empirically calibrated bolometric corrections. 

In short, the synthesis of FASTAR models is based on four main ingredients: stellar templates, a user-defined IMF, a set of isochrones, and a grid of bolometric corrections. We detail below how these four ingredients are treated in the FASTAR synthesis in order to preserve the differentiability of the models and a fast computation time. We also describe the default values for the most relevant parameters in the synthesis process, although they can all be adjusted by the user to generate their models.

\subsection{IMF parameterizations}
The default FASTAR predictions are normalized to one solar mass at birth, i.e., the weighting of the number of stars in Eq.~\ref{eq:1} is calculated by normalizing  $\chi(m)$ between $m_0=0.1 \, \mathrm{M}_{\odot}$ and $m_1= 100 \, \mathrm{M}_{\odot}$ so that 

\begin{equation} \label{eq:2}
    \int_{m_0}^{m_1} 
      \,\chi(m) \, \,m \, dm,
    = 1
\end{equation}

Assuming this normalization, FASTAR comes with the six pre-defined functional forms for the IMF. Their basic properties are as follows: 

\begin{description}
    \item[\textit{Milky Way-like IMFs.}] SSP models with a Milky Way-like IMF can be synthesized assuming either a \citet{mw} or a \citet{Chabrier} functional form. For these IMFs, only the low-mass and high-mass integration limits of Eq.~\ref{eq:2} can be treated as a variable parameter.
    
    \item[\textit{Single power-law.}] This is a generalization of the \citet{Salp:55} IMF\footnote{It is worth emphasizing that E. Salpeter never proposed this functional form across the entire stellar mass domain of Eq.~\ref{eq:2}.} where the main free parameter is the linear slope $\alpha$ which in FASTAR has a default, Salpeter-like value of 2.35. 
    
    \item[\textit{Broken power-law.}] A three-segment IMF parameterization as implemented in the \citet{Conroy12} models. This parameterization is determined by two breaking masses (default  m$_\mathrm{break1} = 0.5$ and m$_\mathrm{break2} = 1.0$) and three slopes connecting them (default $\alpha_1$ = 1.3 representing the low-mass end, $\alpha_2$ = 1.8, and $\alpha_3$ = 2.3 corresponding to the high-mass end slope up to the integration limit). 
        
    \item[\textit{Bimodal IMF.}] The same functional form of the MILES models as defined in \citet{vazdekis96} with the high-mass end slope $\alpha_\mathrm{B}$ as single free parameter with a default value of $\alpha_\mathrm{B}=2.3$. Note that, in contrast to the standard definition of this IMF in the MILES framework, FASTAR describes the IMF in linear units. Thus, the default value for the bimodal IMF mimics a Kroupa-like IMF and corresponds to $\Gamma_\mathrm{B}=1.3$ in the original notation of the MILES models. Our implementation also allows for a variable low-mass end slope $\beta_\mathrm{B}$ (the default value adopted in the MILES models is $\beta_\mathrm{B}=1$).
    
    \item[\textit{Tapered power-law.}] We follow the IMF parameterization presented in \citet{Guido05}, with a peak mass m$_\mathrm{p}$, a high-mass end slope $\alpha$ and a tapering exponent $\beta$ for the low-mass end. This functional form is particularly flexible and naturally captures any potential characteristic mass of the IMF, providing a direct link to theoretical predictions. The default values are $\alpha=\beta=2.3$ and m$_\mathrm{p}=0.5$ \msun.

\end{description}

\begin{figure}
    \centering
    \includegraphics[width=8cm]{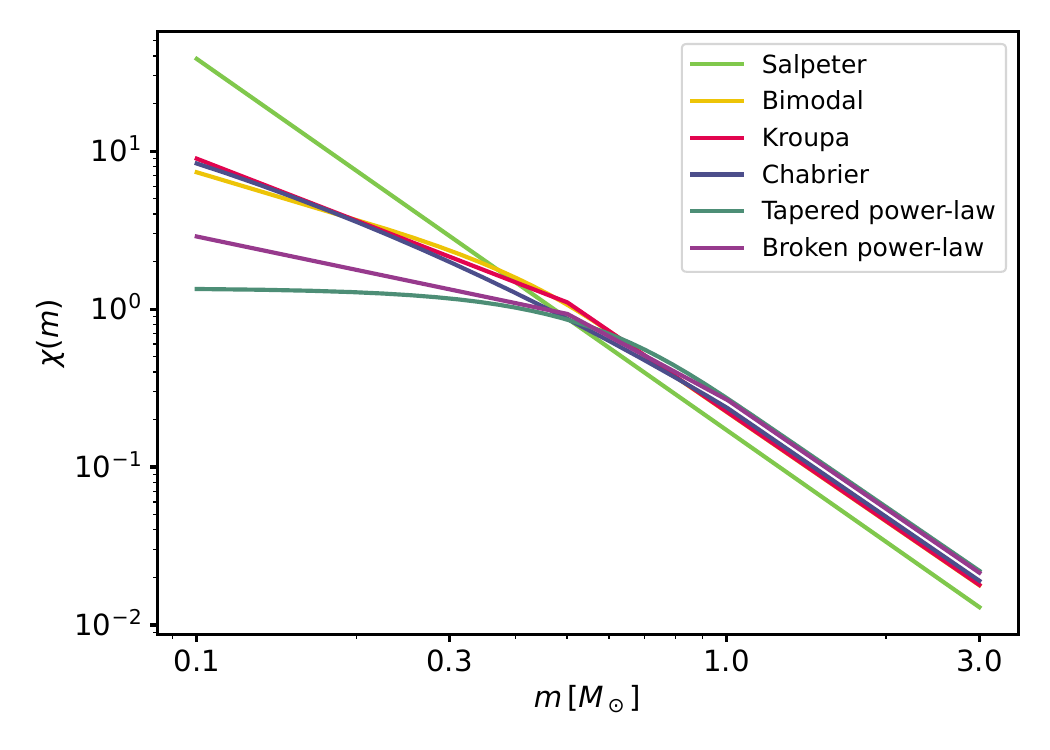}
    \caption{IMF functional forms. FASTAR comes with the six pre-defined IMF parameterizations shown here. All of their parameters can be changed to synthesize new models with variable IMFs. In addition, models for any arbitrary IMF defined by the user can also be calculated.}
    \label{fig:imfs}
\end{figure}

Figure~\ref{fig:imfs} summarizes and compares all these functional forms. In addition, an important asset of FASTAR is that it can also synthesize models with arbitrary, user-defined IMF parameterizations. Examples and guidelines for implementing new IMF functional forms are included in the documentation.

\subsection{Isochrones and bolometric corrections}
FASTAR is based on the most updated version of the BaSTI set of isochrones \citep{Hidalgo18,Pietrinferni21,Pietrinferni24}. Specifically, FASTAR makes use of the solar-scaled predictions, with improved treatment of atomic diffusion and overshooting beyond the Schwarzschild limit. This choice of isochrones in combination with the stellar libraries (see details below) sets the limits of the FASTAR predictions, which are limited to ages between 20 Myr and 14 Gyr and total metallicities between $-2.5 \geq [\mathrm{M}/\mathrm{H}] \leq +0.3$. For an extended discussion of the isochrone behavior we refer the reader to \citet{Hidalgo18}.

It is worth noting that, while the isochrones are calculated for a solar-scaled elemental mixture ([$\alpha$/Fe]=0), individual stellar templates might have non-solar abundances. In practice, this means that in the synthesis of the FASTAR models based on Eq.~\ref{eq:1}, we made the common assumption that the [Fe/H] measured for the individual stars approximately corresponds to their total metallicity [M/H]. Because of that, the FASTAR models should be regards as base models, according to the nomenclature in the MILES models' papers. Details on the implications of this assumption can be found in \citet{miles,Vazdekis15}.

To provide continuous and differentiable SSP predictions requires the interpolation of the isochrone predictions and given the complexity of stellar evolution physics this can be a particularly challenging task \citep{Dotter16}. Thankfully, BaSTI isochrones have been intentionally computed to maintain the same number of stellar masses for any age and chemical composition. This greatly simplifies the synthesis as they can be arranged in a regular grid and then linearly interpolated, as implemented in FASTAR. 

In addition, as noted above, FASTAR makes use of empirical bolometric corrections to project the theoretical isochrone predictions for the bolometric luminosities into  V-band magnitudes directly measurable in the individual stellar templates. FASTAR relies on the V-band bolometric corrections of \citet{Worthey11} for all stellar types and metallicities. These predictions are also shaped into a regular grid and then can be linearly interpolated at any desired value of $T_{\mathrm{eff}}$, $\log g$, and [Fe/H].

\subsection{Stellar libraries}

FASTAR is built around the Medium-resolution Isaac Newton Telescope library of empirical spectra (MILES) \citep{Pat06}. In short, MILES is a compilation of 985 empirical stellar spectra at a spectral resolution of 2.51 \AA \ \citep[][]{Jesus11} covering a wide range of atmospheric parameters (e.g. from $\sim -2.5$ to $\sim+0.5$ in [Fe/H], including templates as hot as $T_{\mathrm{eff}} \sim 25,000$K). With an exquisite flux calibration over the entire $3,540$--$7,400$ \AA \ wavelength range (down to the $\sim2$\% level), MILES feeds many state-of-the-art SSP models currently available.

While the MILES stellar library provides excellent performance for detailed spectroscopic measurements, photometric analyses usually require combining filters that lie blueward and redward of the MILES wavelength limits. Therefore, FASTAR also makes use of the last version of the BOSZ synthetic stellar spectral library presented in \citet{Meszaros24}. This theoretical library has predictions over a much broader wavelength range (from 100 nm to 32 $\mu$m) while simultaneously offering an atmospheric parameter coverage similar to that of MILES (although their predictions are limited to stars cooler than 16,000 K). To be consistent with the FASTAR assumptions, we only make use of the \citet{Meszaros24} predictions for [$\alpha$/Fe]=0 assuming a microturbulent velocity of $v_{micro}=2$ km\,s$^{-1}$. With these additional constraints, our final theoretical library is composed of 6,151 individual stellar spectra. Details on how these two libraries are combined are provided in the following sections.

\subsection{Spectral interpolation}
The interpolation of stellar spectra is arguably the central component of evolutionary stellar population synthesis codes. The goal is to predict the spectrum of a star given its $T_{\mathrm{eff}}$, $\log g$, and [Fe/H] and current state-of-the-art models follow two main interpolation approaches. Local interpolators use stellar templates close to the desired ($T_{\mathrm{eff}}$, $\log g$, [Fe/H]) in order to predict the output spectra \citep[e.g.,][]{vaz03,Sharma16}. On the contrary, global interpolators parametrize the behavior of the stellar library through high-order polynomial fits in order to synthesize new stellar spectra \citep[e.g.,][]{Prugniel11,Villaume17}.

FASTAR implements an alternative, machine learning-powered algorithm that is fast and numerically differentiable. While sophisticated spectral interpolators based on large datasets have been recently explored \citep[e.g.,][]{Melchior23,Koblischke24}, FASTAR follows the simple yet effective approach presented in \citet{Alsing20}. In short, we first reduce the dimensionality of the stellar libraries using a principal component analysis (PCA) decomposition. Then, a fully connected neural network is used to estimate the mapping between the PCA coefficients and the stellar parameters ($T_{\mathrm{eff}}$, $\log g$, [Fe/H]). 

The workflow is as follows. We first convolve the \citet{Meszaros24} stellar library to match the resolution and wavelength sampling of the MILES templates \citep{Jesus11}. Then, we augment ten times each star within the MILES library using a simple linear interpolation. The augmentation process is rather insensitive to the interpolation scheme since we are only perturbing the spectra within the typical uncertainty of the atmospheric parameters, and allows us to bias the solution toward the empirical spectra in those regions where MILES stellar templates are available. 

Equipped with theoretical and the (augmented) empirical templates, we then use a PCA decomposition to reduce the dimensionality of the spectra. In the same line as previous works \citep[e.g.,][]{Yip04,Portillo20,Sharbaf23,Patricia24,Patricia25}, we find that a relatively small number of components is enough to accurately model our stellar templates. After testing a range of values, we settle on 16 principal components for our analysis. Prior to the PCA, each spectrum is first normalized by its own mean flux and then centered by subtracting the mean spectrum of the entire training set. This procedure removes global flux-level differences and provides stable, zero-centered inputs for the PCA decomposition. Finally, we train a fully connected neural network with four hidden layers (64, 128, 128, 64 neurons) using a mean squared error loss to predict the 16 principal component coefficients of the spectra from the stellar atmospheric parameters ($\log g$, $T_{\mathrm{eff}}$, [Fe/H]). During the training, empirical and stellar templates are used indistinguishably to calculate the loss, with no specific merging strategy. Once trained for 1,200 epochs on a standard GPU, the regressor maps stellar parameters to PCA coefficients, which can subsequently be transformed back into full spectra through the PCA basis. Note that, given the limited number of stellar templates available and their sparsity, all of them are used during training.

\begin{figure}
    \centering
    \includegraphics[width=8cm]{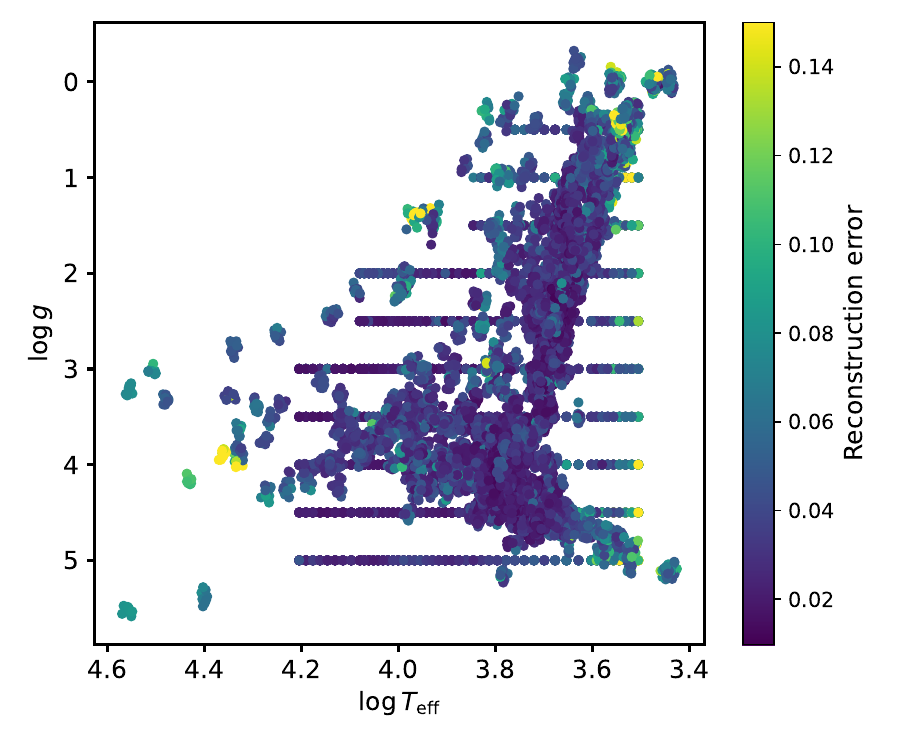}
    \includegraphics[width=8cm]{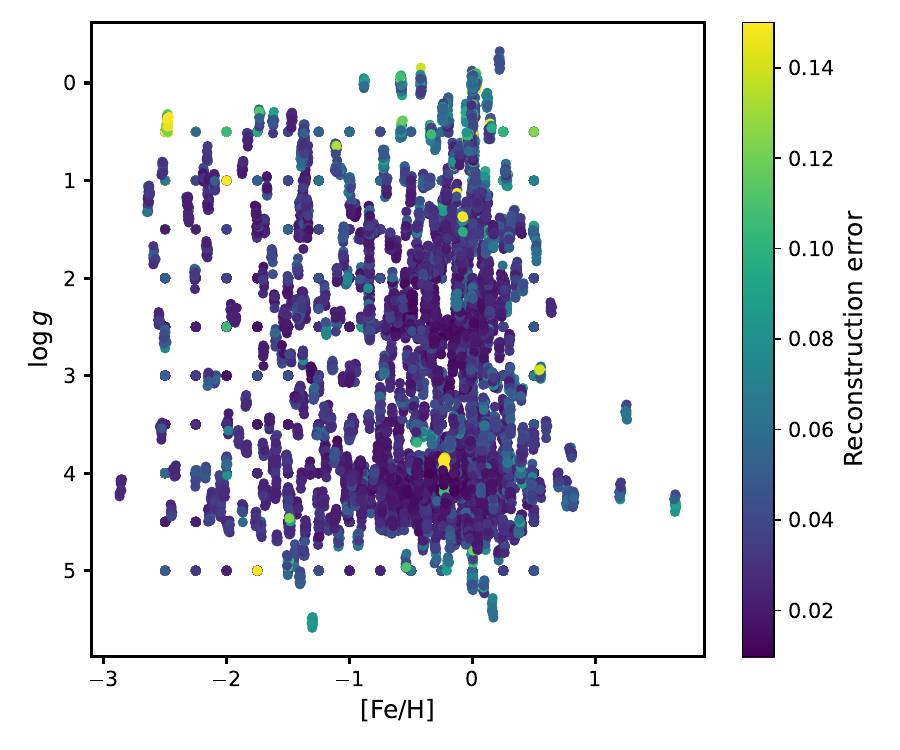}
    \caption{Performance of the FASTAR interpolator across the $\log g$-$T_{\mathrm{eff}}$-[Fe/H] parameter space. Each star in the MILES and BOSZ libraries is color coded by the reconstruction error, defined as the standard deviation of the residuals between the real spectrum and the predicted one. The top panel shows the $\log g$-$T_{\mathrm{eff}}$, while the bottom one shows $\log g$-[Fe/H]. The average error is on the order of $\sim3$\%, only becoming closer to $\sim10$\% for the coolest stars in the sample. There is no significant trend with metallicity.}
    \label{fig:recon}
 \end{figure}

 Figure~\ref{fig:recon} summarizes the performance of the FASTAR spectral interpolation, with stars in the training set color coded by the reconstruction error measured as the standard deviation of the residuals between original and reconstructed spectra. The typical reconstruction error is on the order of a few percent ($\bar{\sigma}=0.03$) and the 99$^\mathrm{th}$ percentile remains around 10\%. Moreover, stellar predictions remain robust across the whole parameter space, only becoming progressively less accurate for the coolest stars. To improve the convergence of the interpolation at these edges, we removed from the training sample theoretical stars cooler than  $\log T_{\mathrm{eff}} = 3.5$, as well as the most extreme dwarf ($\log g > 5$) and giant ($\log g < 0$) stars, whose spectra are dominated by broad molecular lines and therefore prone to model systematics. No additional calibration was needed to combine the spectra and atmospheric parameters of MILES and BOSZ, revealing a robust consistency in the atmospheric of both libraries. Table~\ref{tab:stars} summarizes the final training sets.

\begin{table}[ht] 
\caption{Stellar libraries used for the training.}
\label{tab:stars}
\centering
\begin{tabular}{c c c c c}
\hline\hline
Name & N$_\mathrm{stars}$ & $\log g$ & $\log T_{\mathrm{eff}}$ & [Fe/H] \\
\hline
MILES & 985 (x10) & $-0.2$ / 5.5 & 3.4 / 4.6 & $-2.5$ / $+0.5$  \\
BOSZ  & 4,917  & $0.5$ / 5.0 & 3.5 / 4.3 &  $-2.5$ / $+0.5$ \\
\hline
\end{tabular}
\tablefoot{The limits on the atmospheric parameters are orientative as their distributions can present extended tails (see Fig.~\ref{fig:recon}).}
\end{table}

\begin{figure}
    \centering
    \includegraphics[width=8cm]{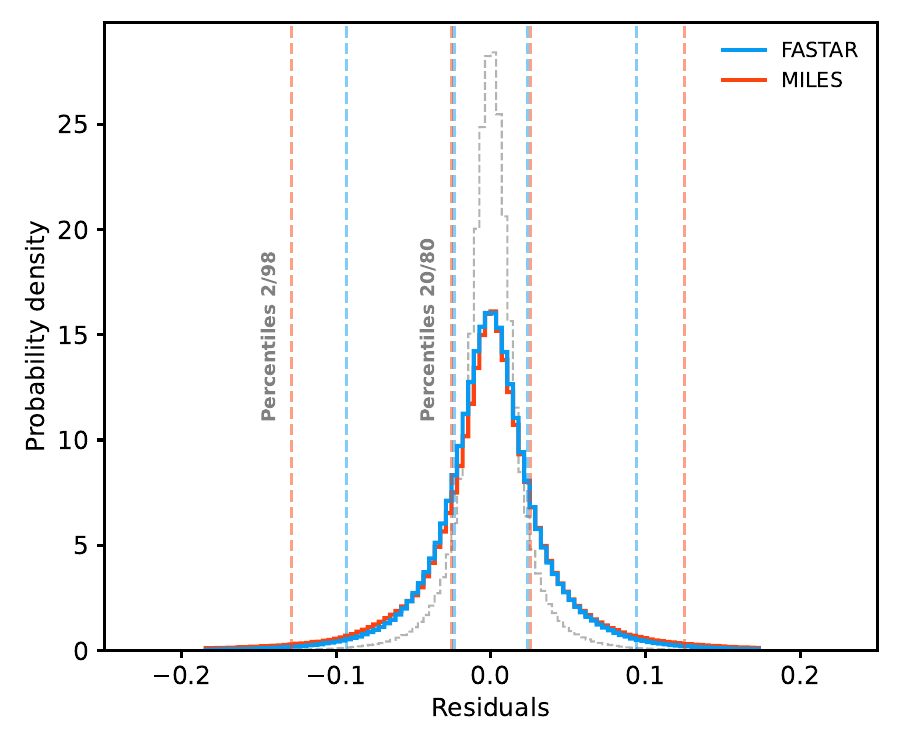}
    \caption{Comparison of residuals. Blue and orange lines represent the distribution of residuals for the FASTAR and MILES interpolators when predicting the stellar spectra of the MILES library. The behavior of both interpolation schemes is similar, as shown by the vertical dashed lines marking the 20$^\mathrm{th}$ and 80$^\mathrm{th}$ percentiles. However, the 2$^\mathrm{nd}$ and 98$^\mathrm{th}$ percentile vertical lines indicate that extreme errors are less common in FASTAR. For comparison, the gray histogram shows the error associated to the PCA compression.}
    \label{fig:resi}
 \end{figure}

Fig.~\ref{fig:resi} shows how the FASTAR interpolator compares with the state-of-the-art local interpolation powering the MILES SSP models. Specifically, it shows the combined residual distribution after predicting the spectra of each star in the MILES stellar library. In general, the behavior of both interpolators is similar although FASTAR residuals exhibit less pronounced tails toward extreme outliers. The error due to the PCA dimensionality reduction is shown in gray. Our tests show that beyond 16 components, the uncertainty in the prediction is dominated by the error associated the neural network training and increasing the number of PCA does not lead to better results.

 \begin{figure*}
    \centering
    \includegraphics[width=18cm]{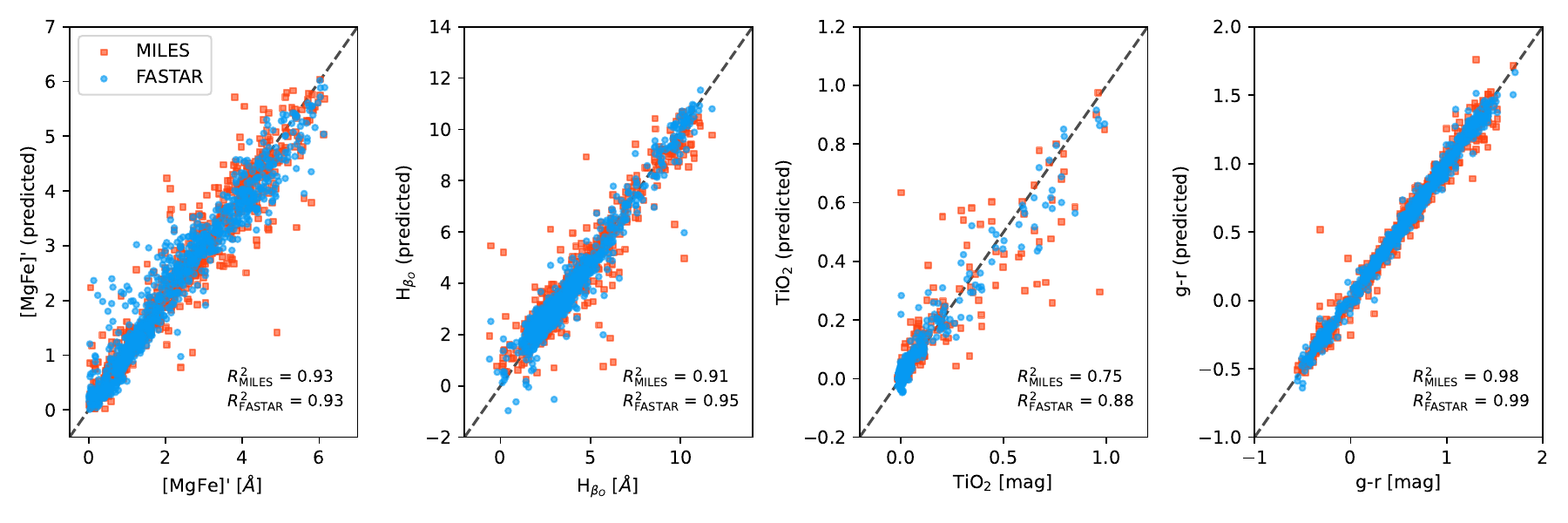}
    \caption{Line-strength and color comparison. Observed (horizontal axes) line-strength and colors are compared to the predicted values (vertical axes) using the FASTAR (blue filled circles) and MILES (orange squares) interpolators. From left to right, we show the behavior of the metallicity-sensitive [MgFe]' \citep{TMB:03}, the age-sensitive H$_{\beta_O}$ \citep{Cervantes}, and the IMF-sensitive TiO$_2$ \citep{Faber85,trager}. The panel on the right shows the comparison of the (g-r) color. The interpolation in FASTAR tends to outperform that of MILES, with systematically smaller residuals as indicated on the bottom right corner of each panel.}
    \label{fig:pred_indices}
 \end{figure*}

Although Fig.~\ref{fig:resi} showcases the overall behavior of the FASTAR interpolation, stellar population properties are encoded in specific spectral regions and scales. To exemplify the reliability of FASTAR in reproducing these features, Fig.~\ref{fig:pred_indices} contains four panels, each of them showing in blue how the FASTAR predictions compare to observed colors and line-strengths in the MILES stellar library. For comparison, Fig.~\ref{fig:pred_indices} also includes the predicted values from the MILES stellar interpolator (orange), as well as the R$^2$ value for each interpolator.

In general, FASTAR performs systematically better than the MILES interpolator at predicting MILES stellar templates, with smaller residuals for all indices and colors and fewer outliers. Only for the [MgFe]' line-strength the performance of both interpolators is similar, with some non-linearity apparent in the FASTAR predictions likely due to the difference in [Mg/Fe] between empirical and theoretical libraries \citep{Milone11,Meszaros24}. In this regard, it is worth noting that FASTAR has been trained to reproduce both MILES and BOSZ libraries, while the MILES interpolator is optimized to predict MILES stars. For completeness, in Appendix~\ref{app:indices} we include a more comprehensive comparison between FASTAR and MILES, including a larger set of indices.

\section{Model synthesis}  \label{sec:resu}

\begin{figure*}
    \centering
    \includegraphics[width=18cm]{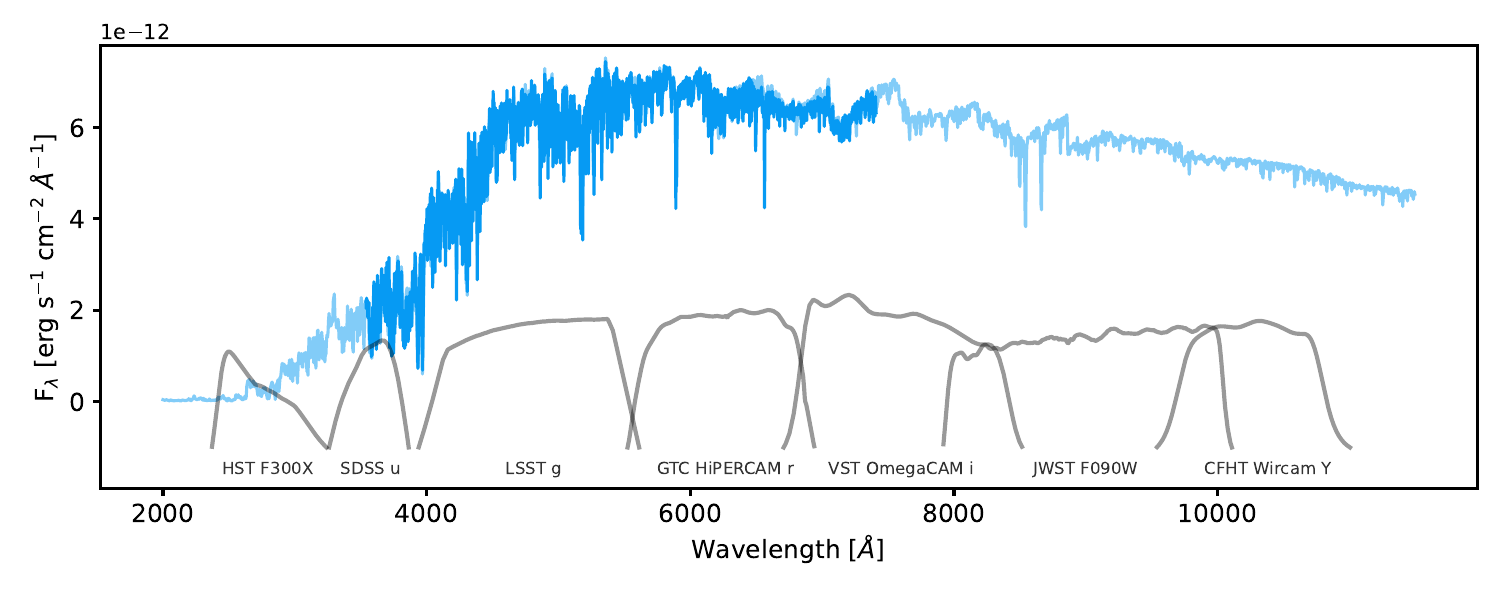}
    \caption{FASTAR spectroscopic and photometric predictions. Blue lines show to the FASTAR SSP models for a 10 Gyr old, solar metallicity population. In light blue, the FASTAR photometric model is based on the BOSZ  theoretical stellar library and can be convolved with any set of photometric filters (exemplified by the different transmission functions in grey) to generate photometric predictions. In addition, the spectroscopic side of FASTAR (dark blue line) combines the MILES (empirical) and BOSZ stellar libraries to generate models oriented toward detailed stellar population analyses.}
    \label{fig:fastar}
\end{figure*}

The evolutionary synthesis of FASTAR is completely defined by the ingredients listed above. The main outcome of FASTAR, i.e., the spectroscopic and photometric predictions are exemplified in Fig.~\ref{fig:fastar} showing the spectral energy distribution expected for an old (10 Gyr), solar metallicity ([M/H]=0) integrated stellar population. The normalization of the FASTAR prediction is anchored to an assumed absolute bolometric magnitude of the Sun of 4.70 and a bolometric correction of BC$_\sun=-0.12$. With this, the units of the predicted fluxes F$_\lambda$ are erg s$^{-1}$ cm$^{-2}$ $\AA^{-1}$, scaled at 10 pc, making  the computation of absolute magnitudes in any photometric band straightforward.

Note that the synthesis of FASTAR, including all its ingredients and operations, is entirely coded using the JAX Python library \citep{jax2018github,jax}, allowing for a optimized computation in both CPU and GPU and native numerical autodifferentiation. The scope of this paper is, however, to describe the evolutionary aspects of the FASTAR synthesis, not its technical implementation. For a more in-depth description of the advantages of JAX in the analysis of stellar populations, we refer the reader to the recent overview presented in \citet{Hearin23}.

FASTAR provides two set of consistent predictions. On the one hand, spectroscopic SSP models can be calculated for any age and metallicity within the isochrone range, combining the MILES empirical stellar library and the theoretical stellar spectra of BOSZ. These spectroscopic FASTAR models are generated at a resolution of 2.51 \AA \ over the MILES wavelength range $3,525$--$7,500$ \AA \ and shown in Fig.~\ref{fig:fastar} as a dark blue line. On the other hand, FASTAR can also be used for photometric analyses over a wider wavelength range ($2,000$--$11,500$ \AA) based solely on the BOSZ theoretical spectra (light blue line in Fig.~\ref{fig:fastar}). This photometric version of FASTAR is at the same resolution as the spectroscopic one but intentionally undersampled ($\Delta \lambda = 4$ \AA) to strongly discourage its use for spectroscopic applications. Note that this wavelength range is narrower than what is in principle allowed by the BOSZ templates, focusing only on those wavelengths commonly covered by optical photometric filters.

\subsection{Line-strengths and colors}

\begin{figure*}
    \centering
    \includegraphics[width=18cm]{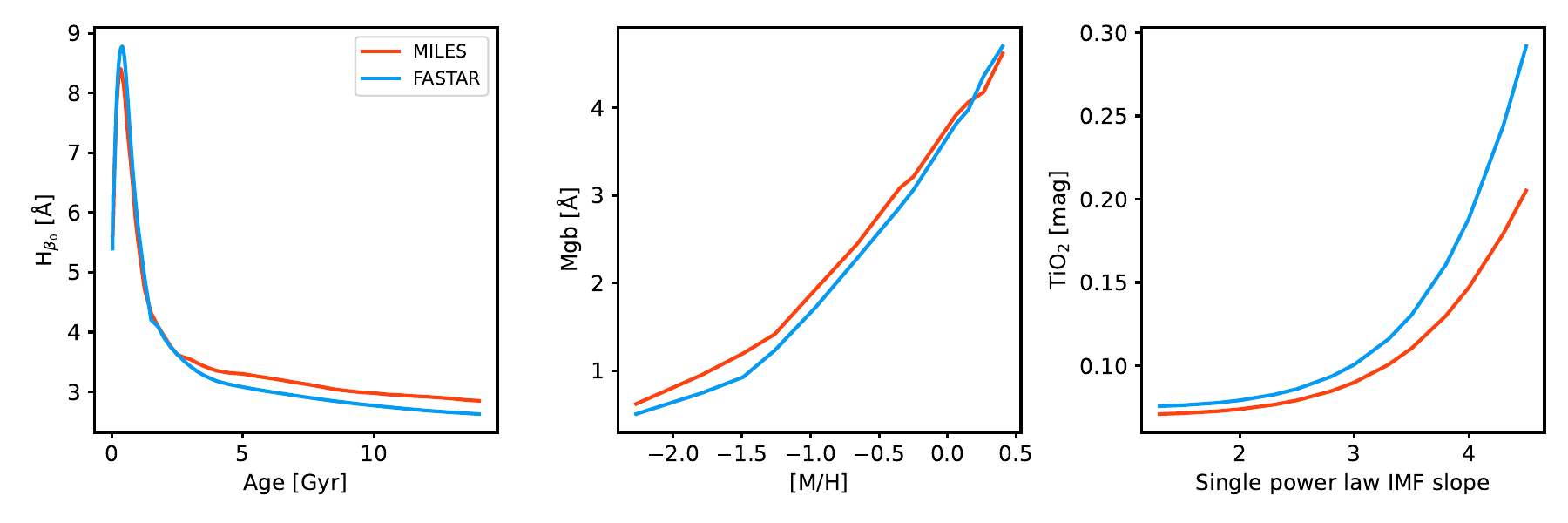}
    \caption{Line-strength comparison. Blue and orange lines indicate, respectively, the FASTAR and MILES predictions for different line-strength indices at a 2.51 \AA \ resolution. Left panel: Equivalent width of the H$_{\beta_O}$ line is shown as a function of age for a population of solar metallicity and a Milky Way-like IMF. For old ages, FASTAR predictions tend to be lower than those of MILES. Middle panel: dependence of the Mgb feature on metallicity for a 10 Gyr population with a Milky Way-like IMF. Right panel: Behavior of the TiO$_2$ molecular band for different IMF slope values (single power law) is shown for 10 Gyr and solar metallicity population.}
    \label{fig:indices}
\end{figure*}

Figure~\ref{fig:indices} presents the FASTAR SSP predictions for the H$_{\beta_O}$, Mgb, and TiO$_2$ as a function of age, metallicity and IMF slope, respectively. For comparison, Fig.~\ref{fig:indices} also includes the MILES predictions for the same set of line-strength indices. In all cases, indices have been measured at the nominal 2.51 \AA \ resolution.

In general, FASTAR predictions follow the expectations for state-of-the-art stellar population synthesis models. Given the similarities in the line-strength measured for the individual stars shown in Fig.~\ref{fig:pred_indices}, the differences between FASTAR and MILES arise from the different choice of isochrones and bolometric corrections. Two important differences are worth mentioning. In the top panel of Fig.~\ref{fig:indices}, FASTAR predicts lower H$_{\beta_O}$ values than MILES for old stellar populations. These low H$_{\beta_O}$ values in FASTAR would tend to alleviate the long-standing zero-point issue of evolutionary stellar population models \citep[see e.g.,][]{Gibson99,Vazdekis01,Schiavon02,Mendel07,Cenarro08,Percival09b,labarbera,Leath22}. Secondly, the bottom panel of Fig.~\ref{fig:indices} shows that FASTAR tends to predict higher TiO$_2$ equivalent widths than MILES for high IMF slope values. This results from the differences in the TiO$_2$ behavior at low effective temperature between the MILES and BOSZ libraries, as demonstrated in Appendix~\ref{app:tio}. It is important to acknowledge that the modeling of these prominent molecular features is particularly challenging, potentially affecting both the theoretical synthesis of stellar templates and the determination of atmospheric parameters in empirical libraries. For completeness, Appendix~\ref{app:index_grid} includes a more extended comparison between FASTAR and MILES for a large set of widely used line-strength indices.

\begin{figure*}
    \centering
    \includegraphics[width=18cm]{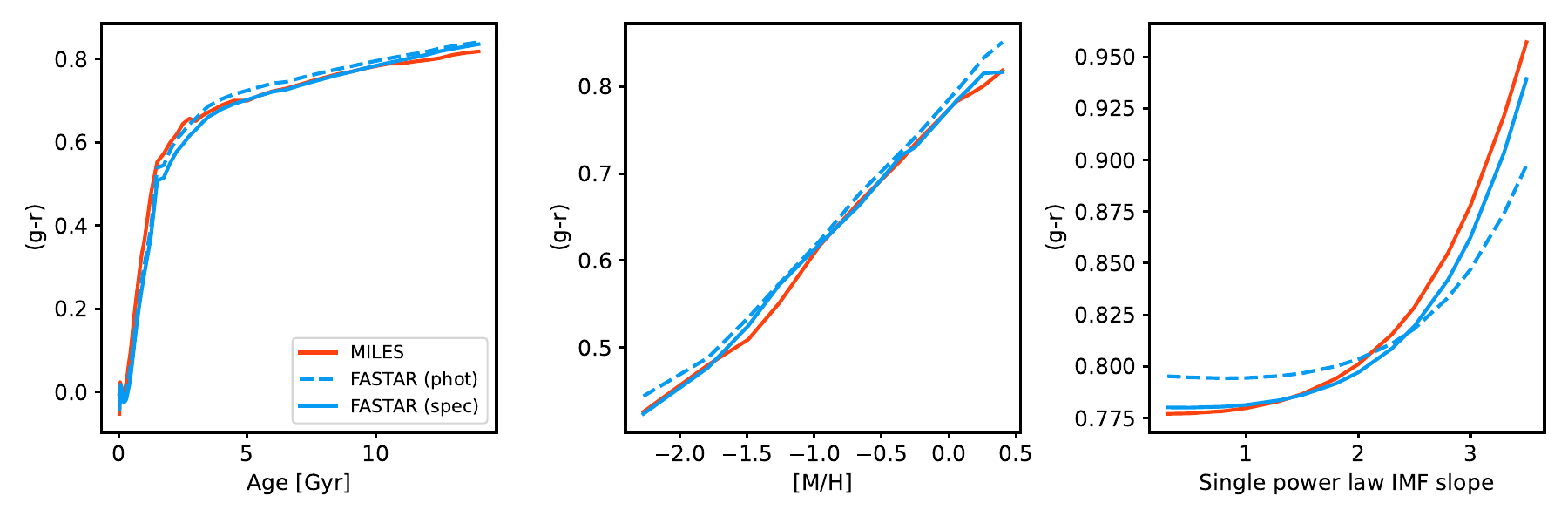}
    \caption{Color predictions. From left to right, each panel shows the color sensitivity of the photometric and spectroscopic FASTAR models (dashed and solid blue lines, respectively), as well as the MILES predictions (in orange) for comparison. The differences between the three models are due to the use of theoretical versus empirical stellar libraries and the choice of bolometric corrections (between FASTAR and MILES). The assumed ages, metallicities and IMF slopes are the same as in Fig.~\ref{fig:indices}.}
    \label{fig:colors}
\end{figure*}

Similarly, in Fig.~\ref{fig:colors} we show the (g-r) color dependence on age, metallicity and IMF slope for FASTAR, using both the photometric and the spectroscopic synthesizer, as well as the comparison with MILES. The three predictions are in overall agreement, with no significant differences between photometric and spectroscopic FASTAR models. The (g-r) dependence on the IMF is also sensitive to the model ingredients. As in Fig.~\ref{fig:indices}, FASTAR tends to show milder changes in color than MILES when increasing the IMF slope value (i.e. the number of low-mass stars). In addition, the photometric FASTAR predictions show an even weaker dependence on the IMF. Given the prominence of broad molecular bands in the atmospheres of cool stars, the differences between the two FASTAR predictions (solid and dashed blue lines in the bottom panel of Fig.~\ref{fig:colors}) are expected to be related to the theoretical versus empirical modelling of these molecular bands.

\subsection{Magnitudes and mass-to-light ratios}

\begin{figure*}
    \centering
    \includegraphics[width=18cm]{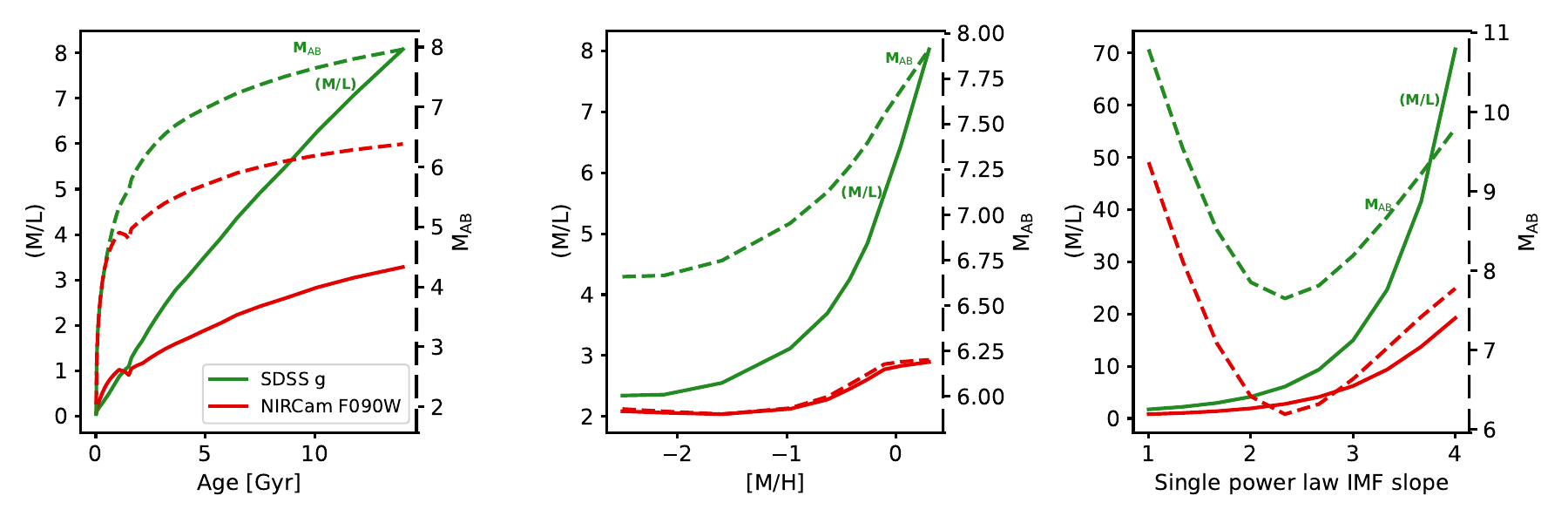}
    \caption{Mass-to-light ratio and magnitude trends. 
    Solid lines and left vertical axes represent the change in stellar M/L as a function of age (left), metallicity (middle), and IMF slope (right). When not varied, the age is fixed to 10 Gyr, the metallicity to solar, and the IMF to the Milky Way value. Similarly, dashed lines and right axes correspond to the expected absolute magnitude of the population. FASTAR predictions for the SDSS g band are shown in green while the expectation for the near-infrared JWST/NIRCam F090W filter are shown in red.}
    \label{fig:ml}
\end{figure*}

Magnitudes and mass-to-light ratios can also be readily calculated within FASTAR. As indicated above, FASTAR predictions are scaled so that the absolute magnitude of any given SSP model is straightforward to compute. In practice, given a generic photometric filter $P$, FASTAR fluxes $F_\lambda$ (from Eq.~\ref{eq:1}) can be translated into absolute AB magnitudes through the following equation:

\begin{equation} \label{eq:3}
     M_{\rm AB}(P) \;=\; -2.5 \,\log_{10}\!\left< F_\lambda \right>_P \;-\; 2.40 \, ,
\end{equation}

with the average flux density $\left< F_\lambda \right>_P$ within the filter given by:

\begin{equation} \label{eq:4}
     \left< F_\lambda \right>_P = \;=\; \frac{\int F_\lambda(\lambda)\,T_P(\lambda)\,\lambda \, d\lambda}.
     {\int T_P(\lambda) / \lambda \,d\lambda} 
\end{equation}

\noindent 
In the equation above, $F_\lambda$ is the predicted FASTAR spectrum, $T_P(\lambda)$ the transmission function of the filter, and $\lambda$ the wavelength array (in \AA). In short, because of the units and scaling of FASTAR, Eq.~\ref{eq:3} predicts the absolute AB magnitude of a stellar population that formed 1 solar mass at birth.

In addition to this, FASTAR can also provide the expected mass-to-light ratio in any photometric filter within the predicted wavelength range ($\sim$3,900 -- 7,000 $\AA$ for spectroscopic FASTAR and 2,000-to-12,000\ $\AA$ for the photometry-oriented models). These mass-to-light ratios are calculated using the CALSPEC Kurucz model of the Sun \citep{Bohlin14}, scaled to our assumed V-band absolute magnitude (see Sect.~\ref{sec:resu}).

FASTAR provides two different mass-to-light ratio M/L predictions. The most directly connected to the stellar population content of galaxies is the stellar M/L. This stellar M/L is defined as the stellar mass of the population (in \msun \ units) divided by its luminosity (also in solar units). However, there are situations where it might be useful to estimate the total mass of a system given its stellar luminosity, including not only the mass in stars but also the mass in stellar remnants and the mass in form of gas ejected by stellar winds. For these cases, FASTAR also provides a total M/L that takes into account all mass sources. If one wants to translate the observed luminosity of a galaxy into a stellar mass, one might use the stellar M/L since the use of the total M/L requires further assumptions (i.e. knowing the behavior of the high mass end of the IMF even if those stars are not present or the assumption that all stellar remnants and gas remain within the observed region or galaxy.)

To illustrate the FASTAR magnitudes and M/L predictions, Fig.~\ref{fig:ml} shows their dependence on the age, metallicity and IMF slope of the underlying population for two different photometric filters, one in the optical and one in the near-infrared. In this case, the mass-to-light ratio corresponds to the stellar ones. Trends are as expected, with higher M/L ratios for older, more metal rich, and bottom-heavier IMF slopes. The brightness of the SSP models increase with decreasing age and metallicities, more significantly in the optical range (green lines). Interestingly, for old ages and solar metallicities, the expected brightness of a population has a minimum for an IMF slope similar to that of the Milky Way. 

\subsection{FASTAR unique features} \label{sec:unique}

\begin{figure}
    \centering
    \includegraphics[width=8cm]{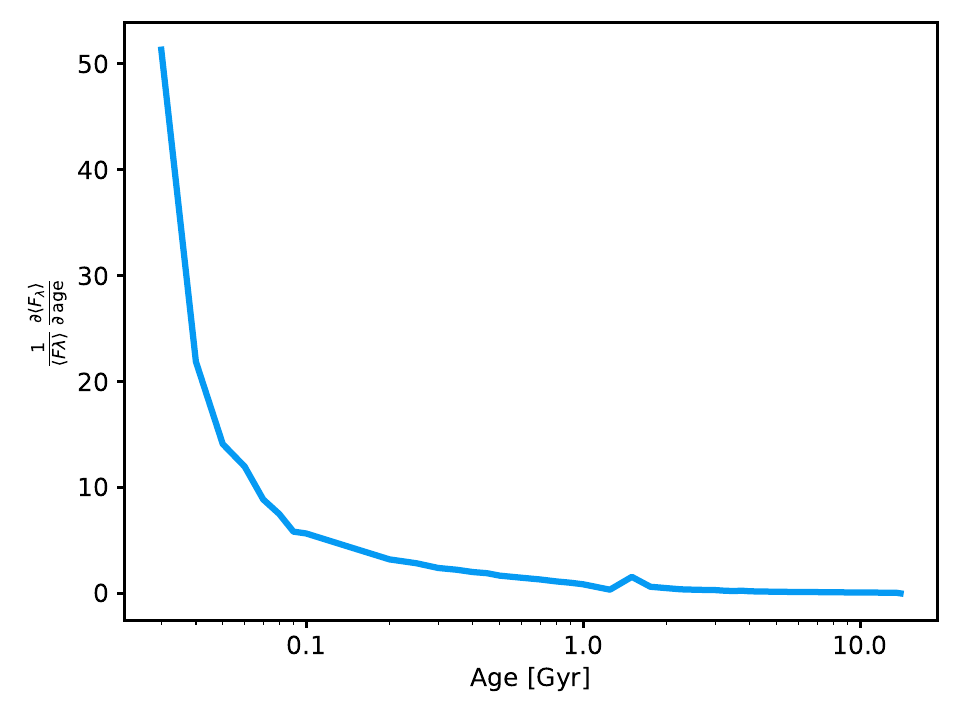}
    \includegraphics[width=8cm]{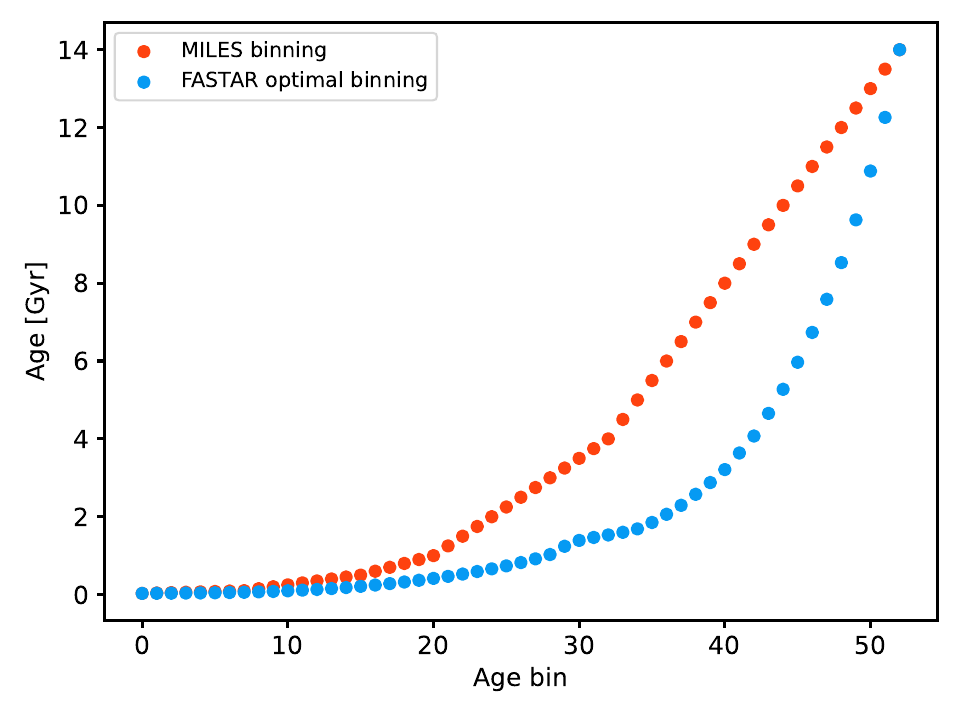}
    \caption{Optimal age sampling. Top: Relative age sensitivity of FASTAR SSP models as a function of age. The normalized age derivative of the FASTAR models becomes smaller for older ages, reflecting the fact that age variations have a progressively weaker effect on the integrated spectrum of a galaxy with increasing age. The peak contribution of the AGB (and RGB) phases appears as a sudden change in the derivative at an age of $\sim 1.5$ Gyr. Bottom: Age sampling of the MILES models is shown in red, following their BaSTI isochrone grid. In blue, the optimized FASTAR sampling enforces a constant flux variation across all age bins.}
    \label{fig:deri}
\end{figure}

\begin{figure}
    \centering
    \includegraphics[width=8.5cm]{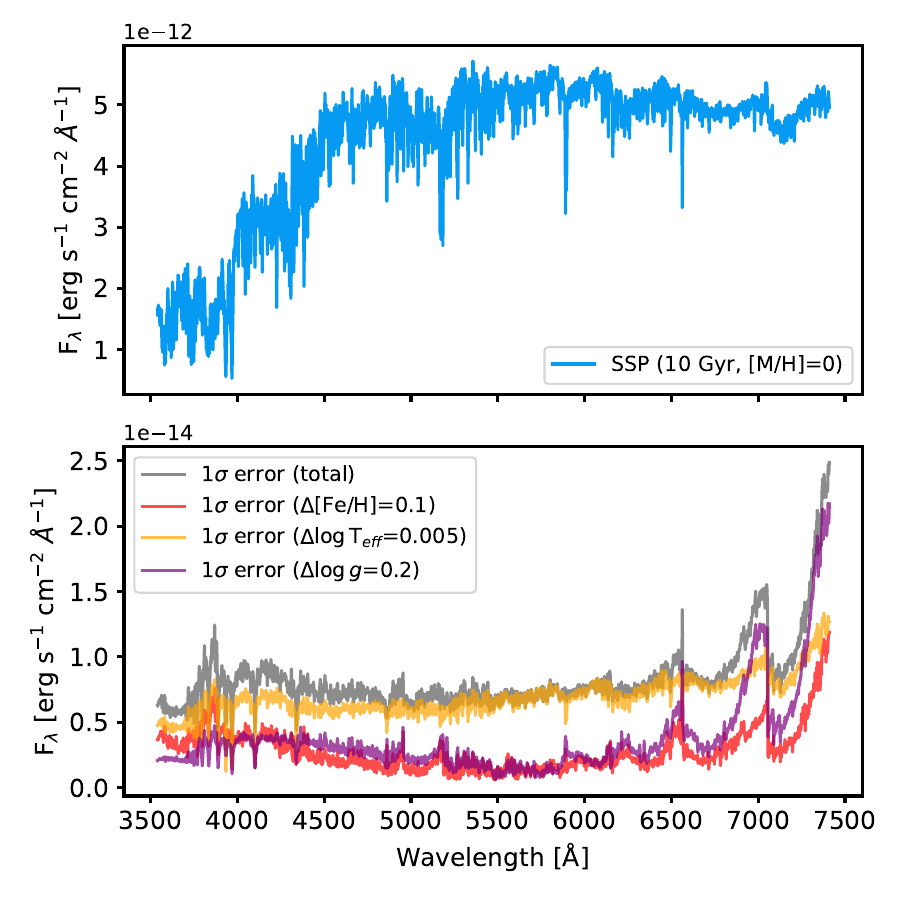}
    \caption{Uncertainty of SSP model predictions. The upper panel shows the SSP model predicted by FASTAR for a population of 10 Gyr and solar metallicity while the bottom panel shows the associated errors. In purple, we present the uncertainty in the SSP model assuming a typical error $\Delta \log g=0.2$ dex in the determination of the surface gravity of the stars. In red and yellow we represent the error corresponding to the metallicity ($\Delta$[Fe/H]$= 0.1$ dex) and effective temperature ($\Delta$T$_{eff}= 0.005$ dex), respectively. In grey, we show the expected total uncertainty in the SSP model.}
    \label{fig:error}
\end{figure}

Spectra and the derived line-strength values, colors and M/L are all standard outputs of evolutionary stellar population synthesis models. With FASTAR, however, it is possible to go beyond these standard predictions. The differentiable and continuous nature of FASTAR can be used, not only to implement more sophisticated gradient-based inversion algorithms, but also to achieve a deeper understanding of SSP models. For example, in the left panel of Fig.~\ref{fig:deri} we show the partial derivative of an SSP model with respect to its age (at solar metallicity). This left panel formalizes the well-known idea that changes in the age have progressively less effect on the integrated spectrum of a galaxy as age increases. In simple terms, distinguishing a 1 Gyr difference is much easier for younger than for older populations. Interestingly, the maximum contribution of AGN and RGB stars to the integrated spectra is also evident as a clear peak in the derivative of the flux at $\sim 1.5$ Gyr.

This robust quantification of the age gradient can be used in practice to improve upon current SSP modelling. Specifically, SSP model predictions are computed in an age grid following the sampling given by the isochrones, as exemplified by the red symbols on the right panel of Fig.~\ref{fig:deri}. This age sampling aims to roughly capture the change in age sensitivity represented in the left panel, with older SSP having a coarser sampling than younger ones. With FASTAR we can formally create an optimal grid of models that ensures that the relative variation of the spectra is the same between two consecutive SSP models, as showcased by the blue symbols on the right panel of Fig.~\ref{fig:deri}. The optimal age sampling also captures the sudden contribution of giant stars to the integrated flux of SSP models.

With FASTAR we provide an optimized age and metallicity sampling based on the mean variation of the spectra. In addition, the sampling can be easily optimized for any other metric (e.g. magnitude in a specific band, lines-strength, color etc.) thanks to the differentiable nature of the model predictions. The documentation released with the code includes examples on how to optimize the sampling for alternative quantities.

In addition, the computation speed of FASTAR can be used to estimate the uncertainty of our spectral and photometric predictions, as illustrated in Fig.~\ref{fig:error}. The top panel shows a standard SSP model of an old (10 Gyr), solar metallicity integrated population. On the bottom panel we show four error spectra calculated as follows. The purple one corresponds to the standard deviation of 100 SSP realizations where the surface gravity of each star along the isochrone was perturbed following a gaussian distribution with dispersion $\Delta \log g=0.2$ dex. In other words, it corresponds to the error in the SSP prediction resulting from the typical uncertainty in the determination of $\log g$ in the individual stellar templates. Similarly, the red line shows the error resulting from propagating a typical stellar metallicity uncertainty of $\Delta$[Fe/H]= 0.1 dex. Finally, with yellow we indicate the uncertainty related to the determination of the effective temperature ($\Delta \log$T$_{eff}=0.005$ dex). The gray line correspond to the total error budget, calculated by simultaneously perturbing the three atmospheric parameters.

With these assumptions, the SSP prediction shown in the upper panel has, on average, a signal-to-noise ratio $\sim 600$ around $\lambda = 5000$, slightly lower ($\sim 300$) toward the blue and the red ends of the optical range. Interestingly, the dominant source of uncertainty across the entire wavelength range is related to the error in the determination of the effective temperature of individual stars. It is worth noting, however, that this is an approximate estimate, assuming constant, uncorrelated and gaussian errors. Moreover, the current version of FASTAR does not take into account the PCA reconstruction loss nor the epistemic uncertainty of the neural network predictions, which will be implemented in upcoming releases. Nonetheless, Fig.~\ref{fig:error} showcases the possibilities of using FASTAR to better understand the strengths and limitations of SSP models.

\subsection{Stellar predictions}

\begin{figure}
    \centering
    \includegraphics[width=8cm]{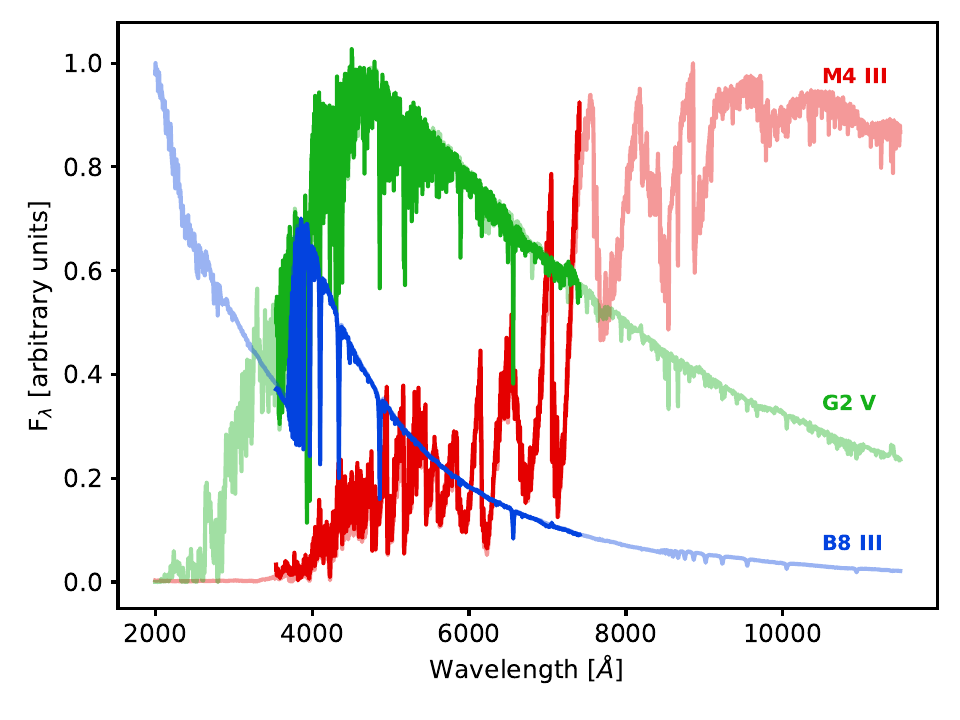}
    \caption{Stellar predictions across different spectral types. Dark and light lines represent predictions using both the MILES and BOSZ stellar libraries, and using only BOSZ, respectively. Each color corresponds to a different stellar type, as indicated by the labels.}
    \label{fig:stars}
\end{figure}

The prediction of stellar templates is implicit to the calculation of SSP models through the neural network-based interpolation scheme described above. To facilitate its usage for scientific cases where individual stellar templates might be required, FASTAR also makes straightforward and fast to predict new stellar spectra. After fixing the metallicity, effective temperature and surface gravity, FASTAR generates stellar predictions across the two (spectroscopic and photometric) wavelength ranges. This is exemplified in Fig.~\ref{fig:stars}.

Darker lines in Fig.~\ref{fig:stars} correspond to the FASTAR predictions within the MILES wavelength range, based on both MILES and BOSZ stellar libraries, while lighter lines are the coarser (in wavelength) FASTAR stellar predictions for photometric applications. Without any fine tuning, the two predictions are in clear agreement. 

\section{Validation on observed data} \label{sec:data}

Globular clusters and massive quiescent early-type galaxies are observational benchmarks for the performance of evolutionary population synthesis models. Their stellar populations are rather simple and homogeneous and thus well-represented by SSP models. 

\begin{figure}
    \centering
    \includegraphics[width=8cm]{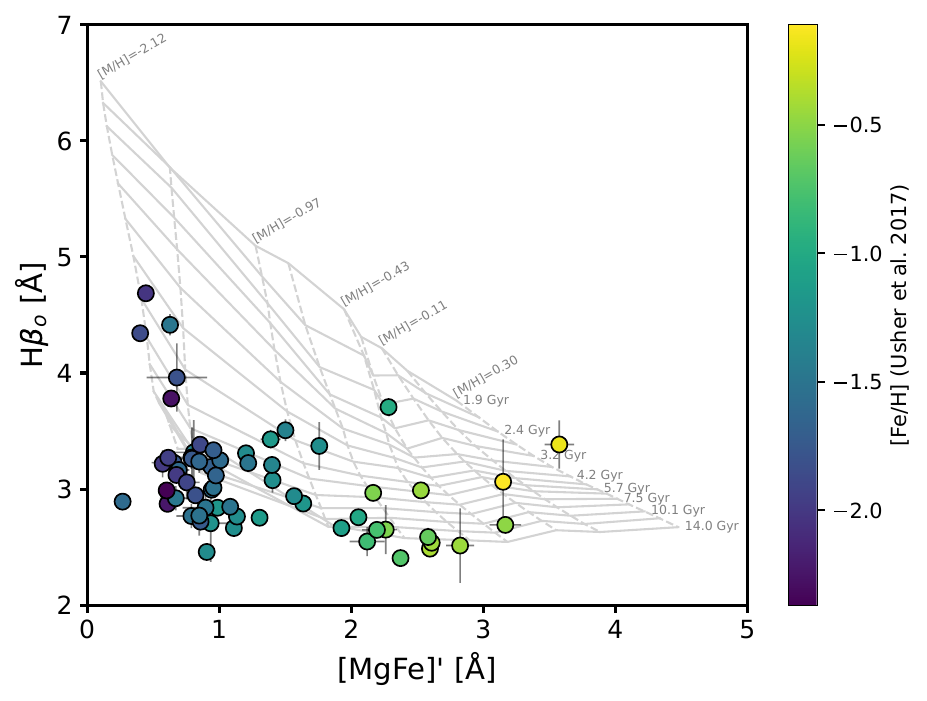}
    \caption{Line-strength H$_{\beta_O}$ versus [MgFe]' index-index diagram for Milky Way globular clusters. Filled symbols correspond to the globular cluster sample of \citet{Usher17}, color coded by their [Fe/H] metallicity. The FASTAR model grid prediction is shown in the background in grey. Labels indicate ages and metallicities across the grid. The equivalent width for data and models is measured at the FASTAR 2.51 \AA \ spectral resolution. FASTAR model predictions reproduce the observed data without systematic offsets, suggesting that the well-known zero-point problem of evolutionary stellar population models can be at least minimized with state-of-the art isochrones.}
    \label{fig:gc_indices}
\end{figure}

In Fig.~\ref{fig:gc_indices} we show the H$_{\beta_O}$ versus [MgFe]' index-index diagram for the Milky Way globular cluster sample of \citet{Usher17}, color coded by their [Fe/H] metallicity. Figure~\ref{fig:gc_indices} also shows the FASTAR predictions for different ages and metallicities, assuming a Milky Way-like IMF. The globular cluster indices were measured at the nominal 2.51\AA \ resolution of the FASTAR models.

In general, FASTAR predictions reliably match the observed data, with old stellar populations across the entire globular cluster population and the independently measured metallicities (i.e. the color code in Fig.~\ref{fig:gc_indices}) running in parallel with the iso-metallicity lines of the FASTAR models. Beyond this agreement, the most notable aspect of this comparison is the lack of a significant zero-point difference between models and data, with the fit to the oldest globular clusters in the sample not requiring ages older than 14 Gyr (i.e. the age of the Universe at $z\sim0$ under standard cosmologies). Interestingly, this well-known systematic of evolutionary synthesis models is present in the MILES models \citep[see e.g.,][]{Leath22} but not in FASTAR, where the predicted H$_{\beta_O}$ values tend to be smaller for the same age (see Fig.~\ref{fig:pred_indices}). Given the similarities between the MILES and FASTAR stellar line-strength predictions (Fig.~\ref{fig:indices}), the apparent lack of discrepancy between FASTAR models and globular cluster data is due to the improved stellar evolutionary physics in the \citet{Hidalgo18} set of isochrones, in particular, thanks to the treatment of atomic diffusion.

\begin{figure}
    \centering
    \includegraphics[width=8cm]{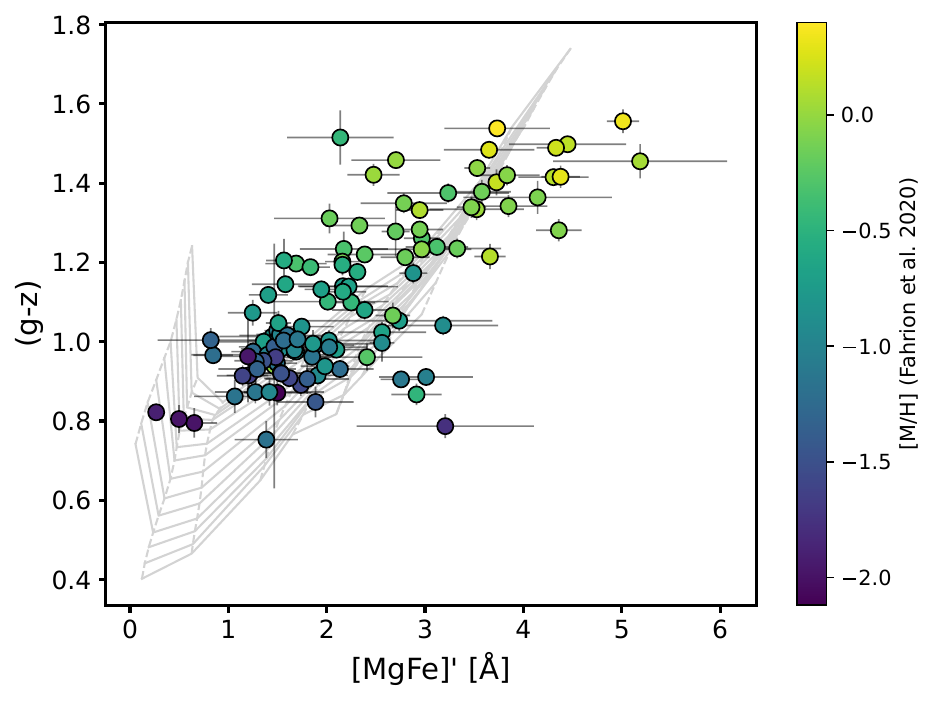}
    \caption{Globular clusters in the Fornax Cluster. Filled circles correspond to the (g-z) versus [MgFe]' values measured in the \citet{Fahrion20} sample of globular clusters in the Fornax galaxy cluster, with the color coding representing their total metallicity. The background model grid shows the FASTAR predictions for different ages and metallicities, with the (g-z) and [MgFe]' values, coming from the FASTAR photometric and spectroscopic sets of predictions, respectively.}
    \label{fig:katja}
\end{figure}

The comparison with observed globular cluster data can be extended beyond the Milky Way. To showcase the broad range of applications of the FASTAR models, Fig.~\ref{fig:katja} shows the (g-z) versus [MgFe]' model grid prediction for a Milky Way-like IMF. On top of the model grid, filled circles correspond to the globular cluster sample of \citet{Fahrion20} in the Fornax cluster, color coded by their total metallicity. Observationally, measuring colors and line-strengths in globular clusters is significantly more difficult in external galaxies than in the Milky Way, in addition to the effect of non-solar elemental abundances, thus explaining the increased scatter in comparison with Fig.~\ref{fig:gc_indices}. However, the agreement between models and data is noteworthy, particularly considering that the color predictions come from the FASTAR photometric models trained on synthetic stellar spectra, while the [MgFe]' line-strengths are measured from the spectroscopic models relying on both theoretical and empirical templates.

\begin{figure}
    \centering
    \includegraphics[width=8cm]{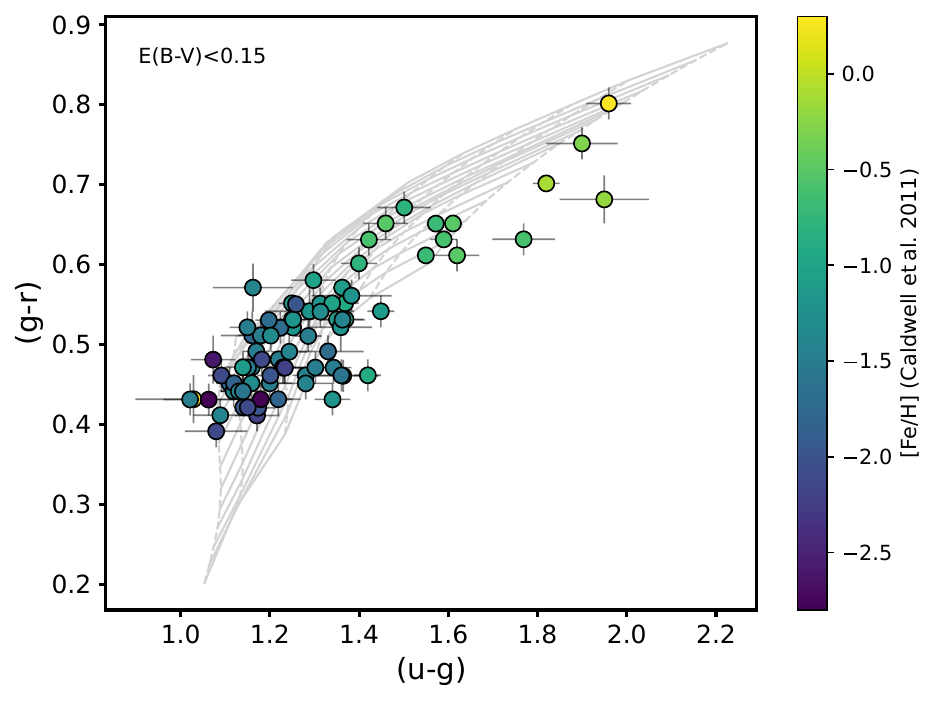}
    \caption{Globular clusters in M\,31. The (u-g) versus (g-r) colors of globular clusters in M\,31 are shown, color coded by their iron metallicity, as reported in \citet{Caldwell11}. To minimize the effect of dust, the sample has been reduced to those globular clusters with E(B$-$V)$< 0.15$. The gray background grid shows the FASTAR photometric predictions for different ages and metallicities, for a fixed Milky Way-like IMF.}
    \label{fig:m31}
\end{figure}

To further benchmark FASTAR predictions against globular cluster measurements, Fig.~\ref{fig:m31} shows the observed (u-g) versus (g-r) colors of M\,31 globular clusters from \citet{Caldwell11}, overplotted on top of a grid of (photometric) FASTAR predictions. The sample of globular clusters has been restricted to those with low attenuation (E(B$-$V)$< 0.15$) to minimize the effect of dust. Clusters are color coded according to their [Fe/H] iron metallicities. The agreement between observations and models is good, with a clear trend between the observed colors, models, and metallicities, in particular for the metal-poor population less affected by dust.

\begin{figure}
    \centering
    \includegraphics[width=8cm]{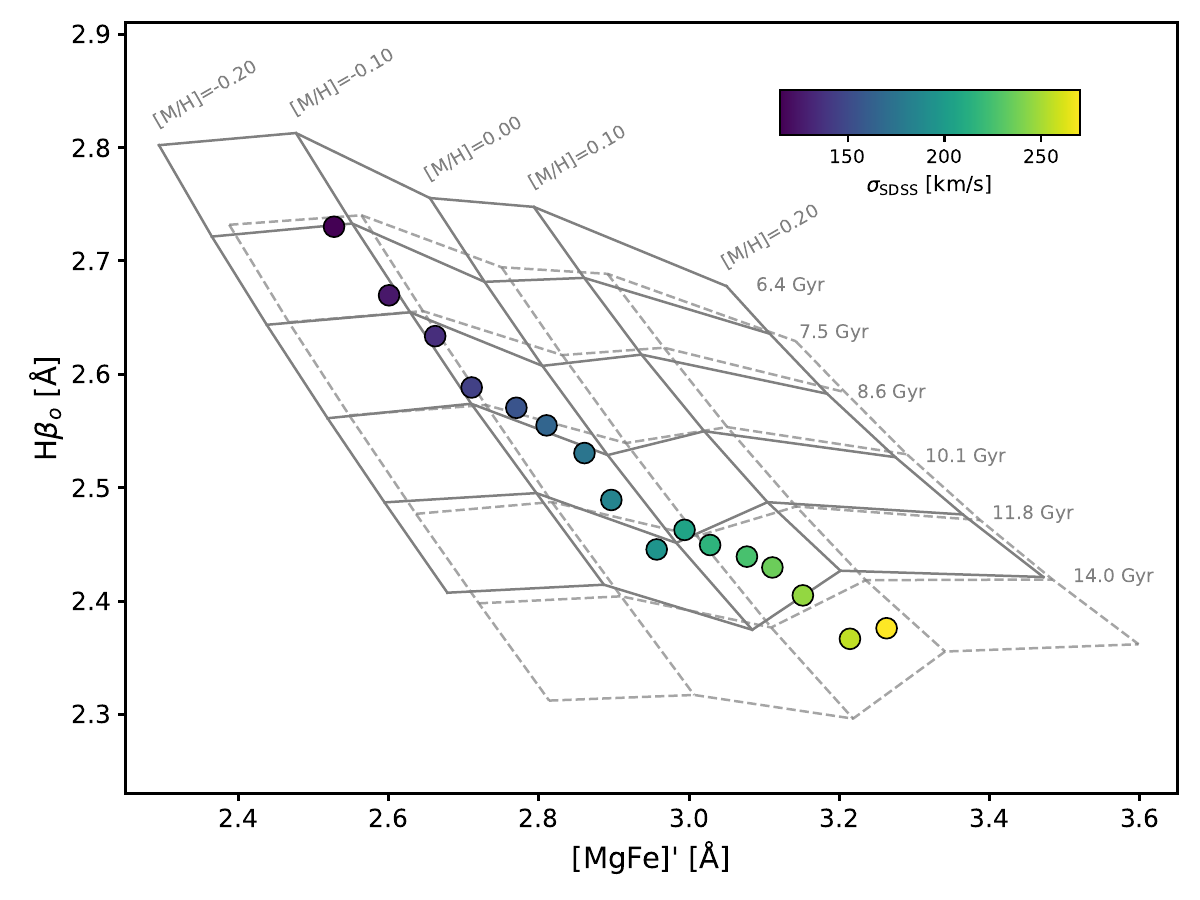}
    \caption{Ages and metallicities of quiescent early-type galaxies. Filled symbols correspond to the H$_{\beta_O}$ and [MgFe]' line-strength measured over the stacked SDSS spectra of \citet{labarbera}, with the color code given by the mean stellar velocity dispersion of each stack. The background model grid shows the FASTAR predictions for those two indices for a Milky Way IMF (solid lines). For reference, the grid for a bottom-heavy IMF (single power law with $\alpha=3$) is also shown with dashed lines. As expected, the prediction from the FASTAR models is that quiescent early-type galaxies become increasingly more metal-rich and older when increasing velocity dispersion. Models and data have been convolved to the resolution of the highest $\sigma$ stack.}
    \label{fig:sdss_indices}
\end{figure}

In addition to globular clusters, we also tested the FASTAR models against the high-quality Sloan Digital Sky Survey \citep[SDSS;][]{DR6} stacked optical spectra of quiescent early-type galaxies presented and described in \citet{labarbera}. Figure~\ref{fig:sdss_indices} shows again the H$_{\beta_O}$--[MgFe]' plane, compared to line-strengths measured from the SDSS stacked spectra, color coded by stellar velocity dispersion $\sigma$. Indices are measured at the spectral resolution given by the stacked spectra with the highest velocity dispersion. As expected, FASTAR models predict a clear age and metallicity dependence with velocity dispersion, with galaxies having higher $\sigma$ presenting older and more metal-rich stellar populations. The recovered comparison is rather similar to the analysis presented in \citet{labarbera} using the MILES models. 

Contrary to globular clusters, the oldest and most massive galaxies in Fig.~\ref{fig:sdss_indices} appear systematically below the H$_{\beta_O}$ prediction for the 14 Gyr SSP. However, this is not necessarily due to a model zero-point issue but to the effect on H$_{\beta_O}$ of a bottom-heavy IMF \citep[i.e., biased toward low-mass stars, ][]{labarbera}, as well as individual elemental abundance variations. For comparison, Fig.~\ref{fig:sdss_indices} also includes, with dashed lines, a FASTAR grid assuming bottom-heavy IMF.

\begin{figure}
    \centering
    \includegraphics[width=8cm]{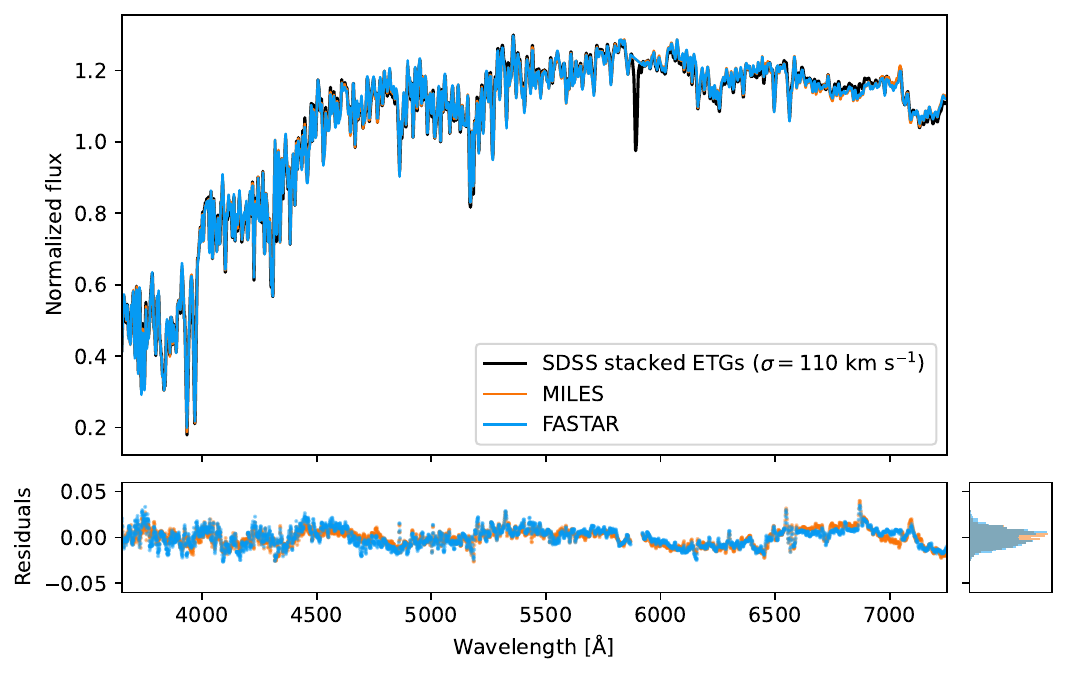}
    \includegraphics[width=8cm]{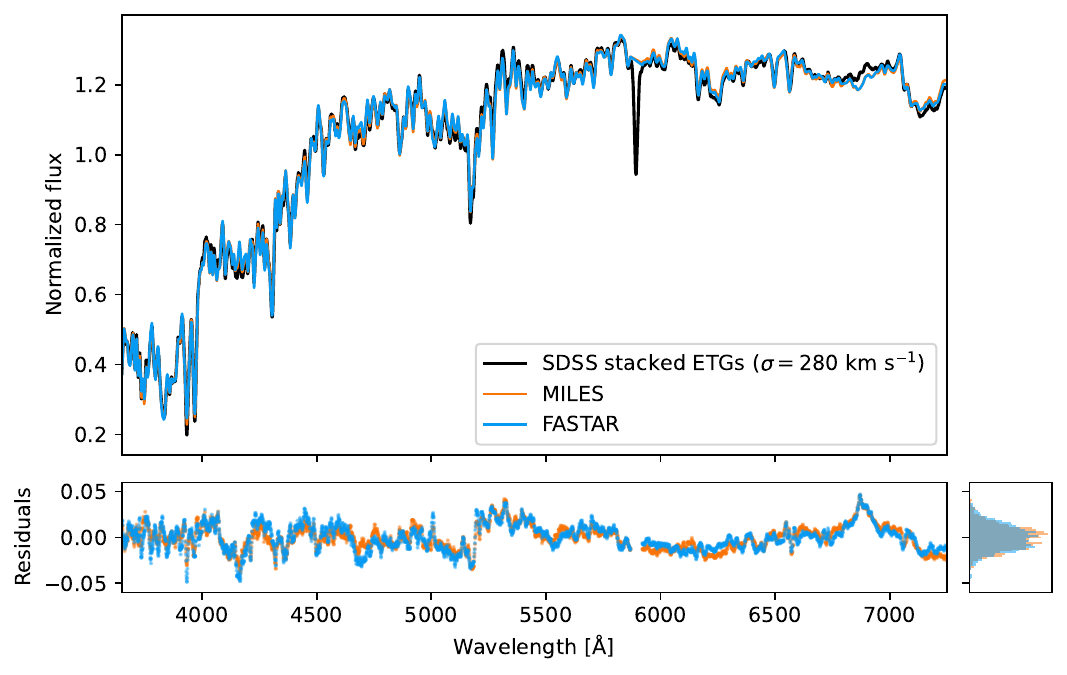}
    \caption{Fits to SDSS stacked spectra. The MILES (in orange) and FASTAR (in blue) best-fitting models are shown compared to two SDSS stacked spectra (in black). The top and bottom panels correspond to stacks of low-$\sigma$ (110 km s$^{-1}$) and  high-$\sigma$ (280 km s$^{-1}$) early-type galaxies, respectively. MILES and FASTAR best-fitting models are very similar as demonstrated by the residual distribution also displayed in each panel. No polynomial correction was applied to the continuum.}
    \label{fig:sdss_stacked}
\end{figure}

Figure~\ref{fig:sdss_stacked} offers a more quantitative assessment of the performance of FASTAR in comparison with the state-of-the-art MILES evolutionary synthesis models. We use pPXF \citep{ppxf,Cappellari23} to fit the SDSS stacked spectra using both sets of models. In the case of FASTAR, we adopt the optimized age-metallicity sampling described in Sect.~\ref{sec:unique}, while fits based on MILES use the default MILES grid. The fits include templates for potential (although weak) emission lines and are done with no continuum correction to avoid compensating potential model systematics. No regularization was applied either, and we masked the NaD region during the fitting process to mitigate the effect of IMF variations and neutral gas absorption.

For both the low- and high-$\sigma$ stacked spectra, the FASTAR and MILES solutions provide very good fits to the data, with residuals below the 5\% level in both cases. This is particularly noteworthy since no polynomial correction is included to account for flux calibration mismatches in the SDSS data. In the case of the high-$\sigma$ stack, the residuals are in both cases slightly larger, likely due to non-solar elemental abundances in the underlying stellar population and a bottom-heavy IMF.

\begin{figure}
    \centering
    \includegraphics[width=8cm]{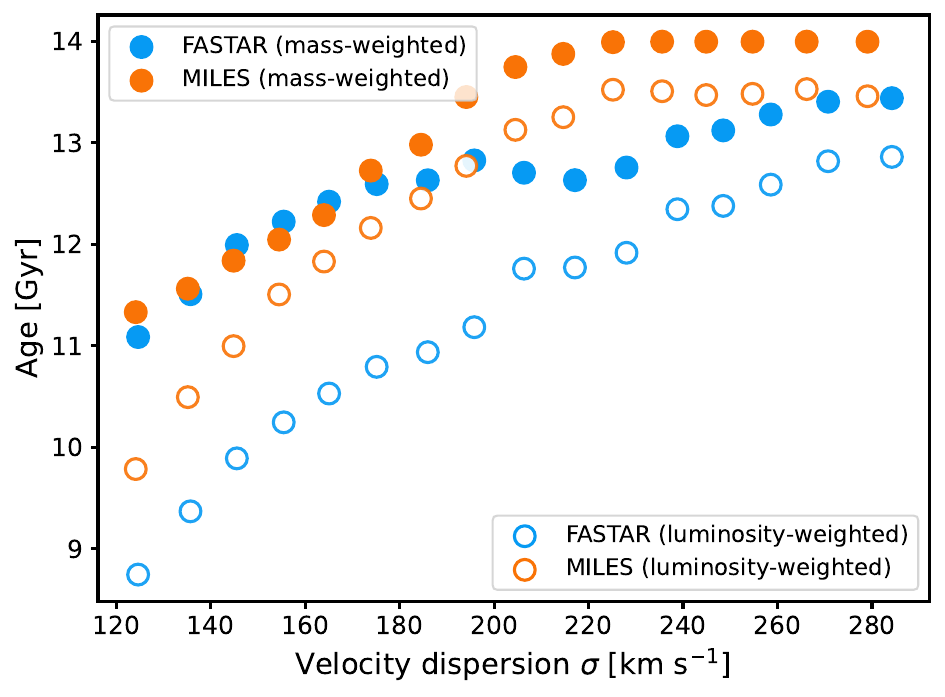}
    \caption{Age-velocity dispersion relation. Blue and orange symbols correspond to the best-fitting ages of the SDSS stacked spectra using pPXF and the FASTAR and MILES SSP models, respectively, while filled and empty symbols indicate mass- and luminosity-weighted values. FASTAR and MILES models are able to retrieve the expected age-$\sigma$ relation, with FASTAR-based ages generally being younger than those measured with MILES.}
    \label{fig:sdss_age_trend}
\end{figure}

\begin{figure}
    \centering
    \includegraphics[width=8cm]{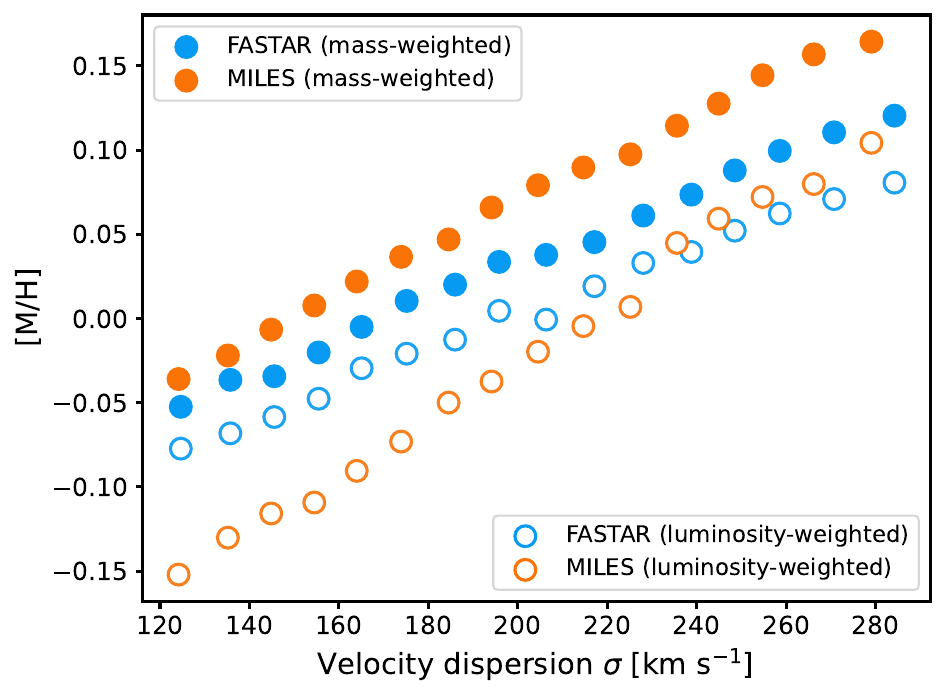}
    \caption{Metallicity-velocity dispersion relation. The symbols and colors are the same as in Fig.~\ref{fig:sdss_age_trend}, but in this case they represent the best-fitting metallicity values. The expected relation with velocity dispersion is well recovered with both sets of models although with systematic differences as FASTAR tends to result in lower metallicities and less differences between mass-weighted and luminosity-weighted values than MILES.
    }
    \label{fig:sdss_met_trend}
\end{figure}

In Figs.~\ref{fig:sdss_age_trend} and ~\ref{fig:sdss_met_trend} we show the recovered age - velocity dispersion and metallicity - velocity dispersion relations, respectively. As before, blue symbols correspond to measurements obtained using the FASTAR models and orange ones correspond to MILES. In both figures, filled symbols are mass-weighted quantities while empty symbols represent luminosity-weighted averages. Generally, FASTAR and MILES models behave similarly, retrieving the well-known trend of older and more metal-rich stellar populations with increasing galaxy velocity dispersion. 

FASTAR tends to retrieve younger ages (Fig.~\ref{fig:sdss_age_trend}), as expected from the line-strength behavior detailed above, with weaker H$_{\beta_O}$ values at fixed age than in MILES. Interestingly, age-$\sigma$ relation measured with the FASTAR model shows a peculiar behavior at $\sigma\sim200$ km s$^{-1}$, which is also evident in the  H$_{\beta_O}$--[MgFe]' plane shown in Fig.~\ref{fig:sdss_indices} but apparently absent when using the MILES models. Regarding the recovered metallicities, FASTAR-based measurements show smaller differences between luminosity- and mass-weighted trends compared to MILES, with systematically lower mass-weighted values.

\section{Summary}\label{sec:summary}

With FASTAR we present the first generation of continuous, differentiable evolutionary stellar population synthesis algorithms. FASTAR is not a set of pre-computed model predictions, but a flexible tool that provides continuous SSP predictions across a wide range of ages, metallicities, IMFs and wavelengths. Based on the fundamental principles of evolutionary stellar population models, FASTAR, however, is faster and enables a deeper understanding of the models' behavior and new ways of using their thoroughly tested strengths. The access to the synthesis code is transparent and allows users to tailor FASTAR predictions to their needs. The key features of FASTAR are as follows:

\begin{itemize}
    \item PCA-based, neural network stellar interpolation. FASTAR is centered around a stellar interpolator trained on empirical and/or theoretical stellar libraries that matches the reliability of competing interpolation schemes currently implemented in SSP models, but it is faster and matches the performance of state-of-the-art alternatives in accurately reproducing stellar spectra.
    
    \item Differentiability and flexibility. FASTAR uses the JAX numerical library to provide differentiable SSP predictions, ready to be implemented in gradient descent inference algorithms. Moreover, it allows for the use of any IMF functional form, facilitating the cross-IMF conversions of colors, stellar masses and all other model predictions.

    \item Age and metallicity coverage. FASTAR relies on the latest version of the BaSTI-IAC set of isochrones, thus enabling SSP predictions from 20 Myr to 14 Gyr in age and from -2.5 to +0.3 in total metallicity. Possible IMF slopes and functional forms are completely unconstrained.
    
    \item Detailed spectroscopy. Over the 3,5400 - 7,400\AA \ wavelength range, FASTAR offers detailed stellar and SSP predictions at a 2.51\AA \ resolution. The code combines the MILES (empirical) and BOSZ (theoretical) stellar libraries to deliver detailed SSP predictions. Colors, magnitudes, mass-to-light ratios, line-strengths etc. can be easily computed as well. 
    
    \item Photometric predictions. For photometric applications, FASTAR offers a wider wavelength range, from 2,000\AA \ to 12,000\AA \ using the BOSZ theoretical stellar library. The resulting SEDs can then be convolved with any photometric filter and the corresponding colors and mass-to-light ratios can be consistently estimated. 
     
    \item Unique to FASTAR. FASTAR proposes an alternative way of releasing SSP models. Predictions can be evaluated at any age, metallicity or IMF. In addition, FASTAR provides an optimal age and metallicity grid sampling based on the formal flux derivative of the models. FASTAR can also be used to assess the intrinsic uncertainty of SSP predictions.
    
\end{itemize}

\begin{acknowledgements}
We would like to thank the referee for the suggestions and comments on the original manuscript. We acknowledge support from grant PID2022-140869NB-I00 from the Spanish Ministry of Science and Innovation. F.L.B. acknowledges support from INAF minigrant 1.05.23.04.01. We would like to thank Ignacio Ferreras, Sebastián Hidalgo, Ignacio Trujillo and Francesco Belfiore for their insightful comments during the development of FASTAR.
\end{acknowledgements}

\section*{Models availability}
Documentation and examples on how to use FASTAR can be found at the project's website
\url{https://fastar.readthedocs.io}

The code is freely available here
\url{https://github.com/inavarro/fastar}

\bibliographystyle{aa} 
\bibliography{fastar_ssp}  

\begin{appendix}

\section{Line-strength comparison} \label{app:indices}

\begin{figure*}
    \centering
    \includegraphics[width=16cm]{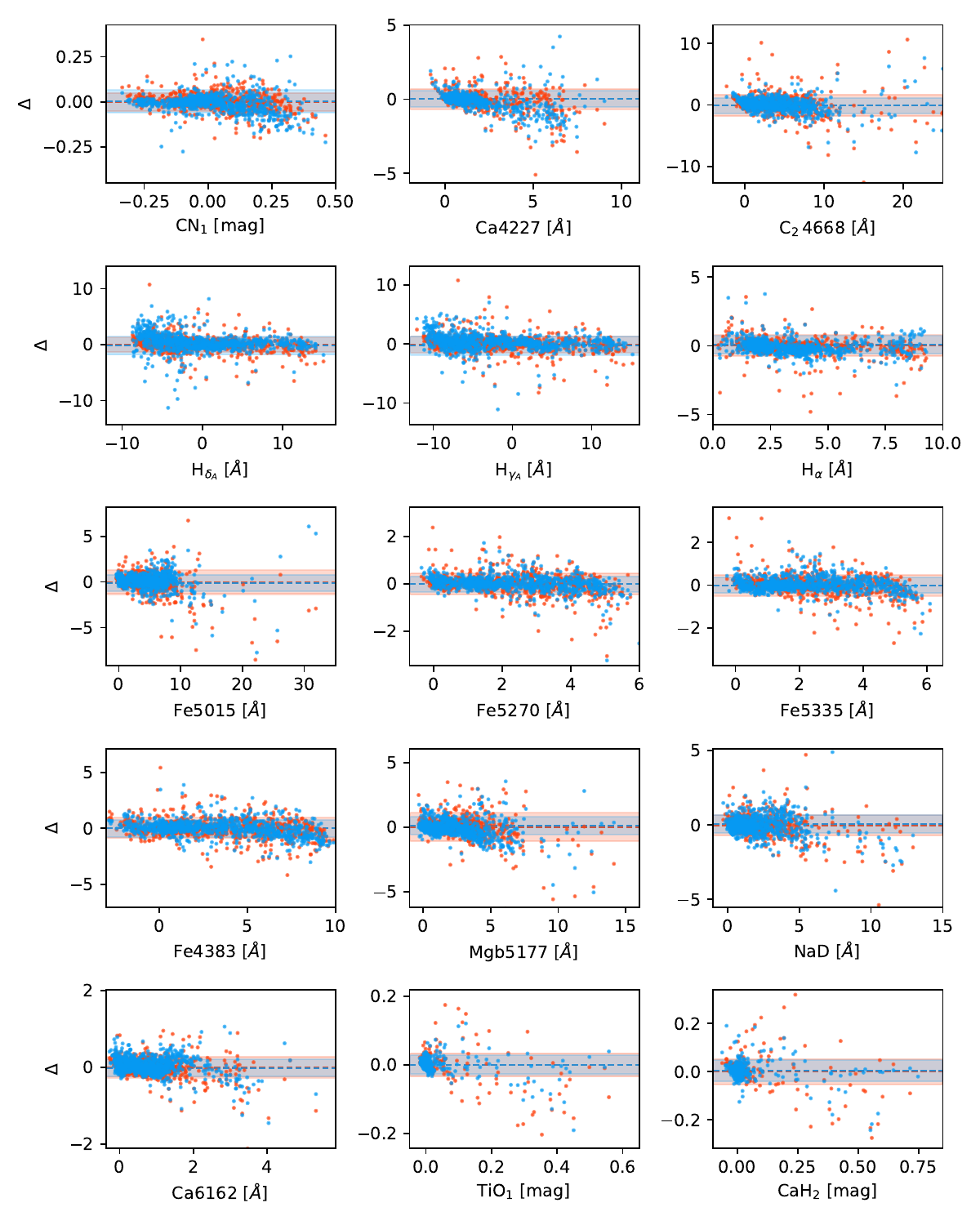}
    \caption{Residuals across the MILES stellar library. Blue symbols correspond to the difference between the FASTAR line-strengths predictions and the actual measurements over the MILES empirical spectra. For comparison, orange symbols indicate the predictions from the MILES interpolator. Dashed lines correspond to the mean of the distribution and shaded areas to the 1$\sigma$ scatter.}
    \label{fig:all_indices}
\end{figure*}

To better assess the performance of the FASTAR interpolation we also charaterize the behavior of different line-strengths indices across the entire spectroscopic range ($3,540$--$7,400$ $\AA$), as shown in Fig.~\ref{fig:all_indices}. Blue symbols correspond to the difference between the predicted value from FASTAR and that directly measured over the MILES stars. We also include a comparison with the MILES interpolation predictions for the same set of indices, in orange. Horizontal dashed lines in Fig.~\ref{fig:all_indices} indicate the mean difference between the observed and predicted line-strength values while shaded areas indicate the standard deviation of the distribution.

As already showcased in Fig.~\ref{fig:pred_indices}, both interpolation schemes perform at a similar level, with FASTAR yielding generally a slightly smaller scatter. Some features such as the Mgb\,5177 or the Ca\,4227 show a small but noticeable non-linearity in the residual distribution due to differences between the empirical MILES and theoretical BOSZ predictions used for training. Yet, in general, the agreement between both interpolators is remarkable, even in features such as the Balmer lines that traditionally pose problems when comparing theoretical and empirical predictions due to non-LTE effects \citep{Knowles23}. In FASTAR, predictions tend to be close to the empirical expectations because the training has been intentionally biased toward MILES through the augmentation process.

It is also worth emphasizing that the FASTAR interpolation has been trained to match both theoretical and empirical libraries, while Fig.~\ref{fig:all_indices} only reflects its performance compared to the empirical stars of MILES. In other words, FASTAR is able to incorporate information about the theoretical BOSZ library while predicting empirical data at the same level as the MILES interpolation. Moreover, the training loss is evaluated based on the PCA components, not, for example, on the line-strength behavior. Then, it would be straightforward to optimize these predictions by explicitly training to also match line-strength predictions, which will be explored in upcoming releases.

\section{TiO$_2$ behavior} \label{app:tio}

The predicted TiO$_2$ values for FASTAR SSP models present a significant systematic deviation with respect to the MILES SSP models, as shown in Fig.~\ref{fig:indices}. Following Fig.~\ref{fig:tio_app}, we interpret this as a consequence of the different behavior of the index with metallicity for stars cooler than $\log T_{eff}\sim3.6$ in the MILES and BOSZ libraries.

It is evident from this figure that, in general, MILES and BOSZ predictions are similar, but MILES stars do not present a significant dependence on metallicity, while stars in the theoretical BOSZ do. However, as noted in the main text, the modeling of broad molecular features can be particularly problematic and prone to systematics whose effects can propagate to both empirical and theoretical stellar libraries. Moreover, the range of metallicities covered by MILES for cool stars is also significantly narrower than that probed by BOSZ.

\begin{figure}[H]
    \centering
    \includegraphics[width=8cm]{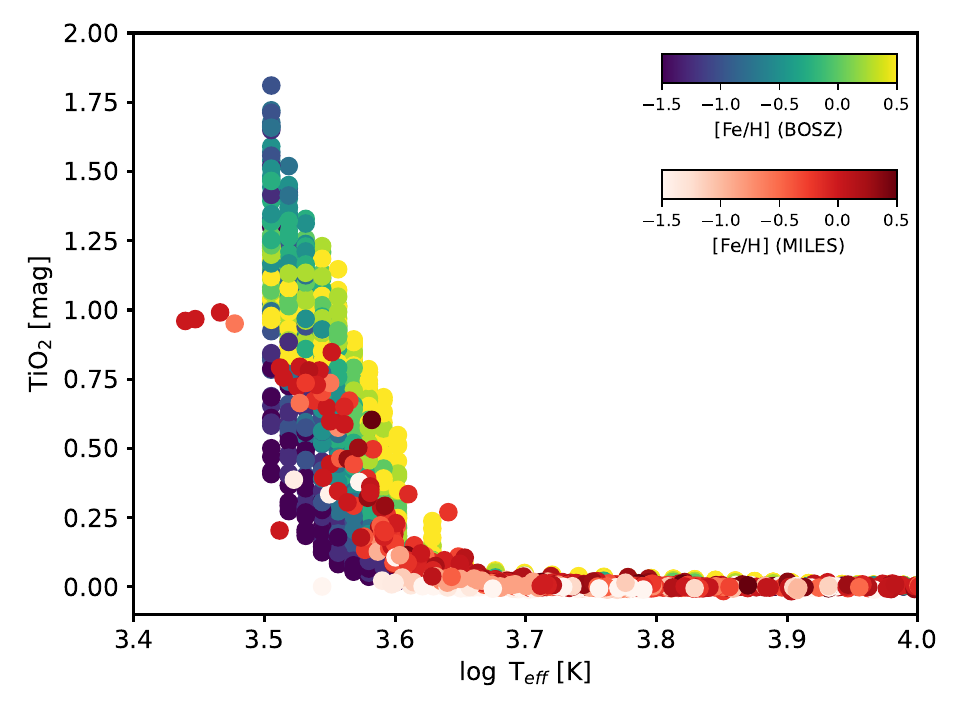}
    \caption{Empirical versus theoretical TiO$_2$ behavior. MILES stars (red symbols) do not exhibit a significant trend with metallicity at low effective temperatures, contrary to the BOSZ library. This explains the differences between FASTAR and MILES in the TiO$_2$ values predicted for SSP models \ref{fig:indices}.}
    \label{fig:tio_app}
\end{figure}

\section{SSP predictions for line-strength indices} \label{app:index_grid}

In the main text we showcase the behavior of FASTAR and MILES SSP predictions for three indices,  H$_{\beta_O}$, Mgb, and TiO$_2$. While that comparison shows an overall good agreement between both sets of models, as well as some differences, Fig.~\ref{fig:indices} does not fully characterize the behavior of FASTAR SSP models. 

\begin{figure*}
    \centering
    \includegraphics[width=16cm]{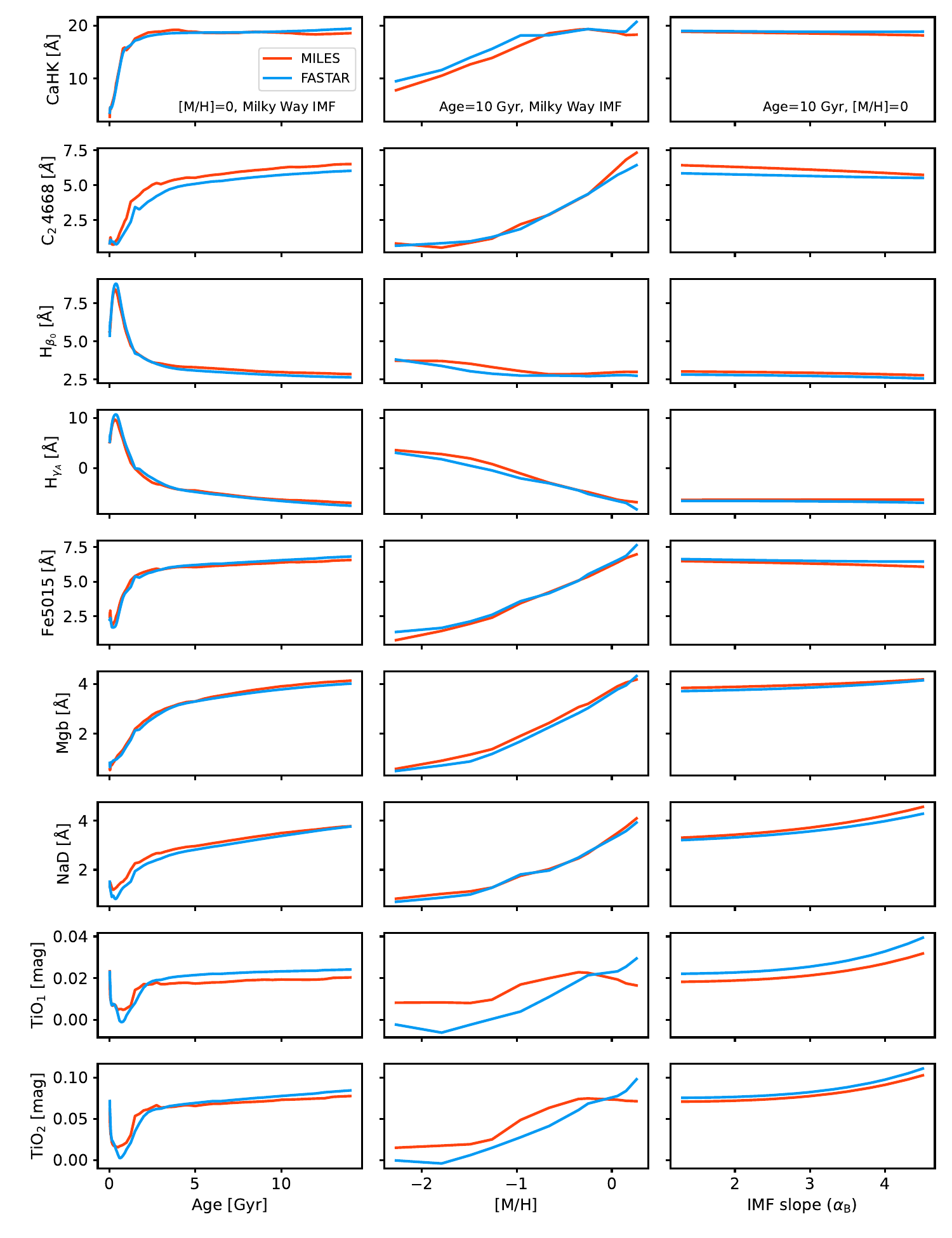}
    \caption{FASTAR versus MILES SSP predictions. Each row shows the dependence of a specific line-strength index with age (left), metallicity (middle) and IMF slope (right, under a bimodal parameterization) for FASTAR (in blue) and MILES (orange) SSP models. In each column, the value of the assumed age, metallicity and IMF slope is indicated in the first row (corresponding to the CaHK feature). Equivalent widths are calculated at the 2.51 \AA \ nominal resolution of both sets of models.
    }
    \label{fig:all_indices_ssp}
\end{figure*}

In Fig.~\ref{fig:all_indices_ssp} we include a larger set of indices, contrasting the FASTAR (in blue) and MILES predictions (in orange) for varying age, metallicity and IMF slope (under a bimodal parameterization in this case). Despite the differences in ingredients and assumptions, Fig.~\ref{fig:all_indices_ssp} shows a remarkable agreement between FASTAR and MILES. From different Balmer lines to features sensitive to metallicity and IMF, the trends and even absolute values predicted by both sets of models are notably similar. Although model systematics are, in general, a limiting factor of SSP modelling, the similarities between FASTAR and MILES shown in Fig.~\ref{fig:all_indices_ssp} clearly speaks in favor of the consistency and robustness of evolutionary stellar population models.

As expected, the largest systematic differences between FASTAR and MILES happen at the extremes of the metallicity range. For example, for the CaHK feature (using the \citealt{serven} definition), the agreement increases for those metallicities around the solar value where the importance of MILES empirical templates is maximal. For the case of the TiO$_1$ and TiO$_2$ features, the differences in the metallicity dependence between MILES and BOSZ (see Appendix~\ref{app:tio}) become evident at the SSP level.

\end{appendix}

\end{document}